\documentclass[sn-mathphys,Numbered]{sn-jnl}

\usepackage{multirow}%
\usepackage{mathrsfs}%
\usepackage[title]{appendix}%
\usepackage{xcolor}%
\usepackage{textcomp}%
\usepackage{manyfoot}%
\usepackage{booktabs}%
\usepackage{algorithm}%
\usepackage{algorithmicx}%
\usepackage{algpseudocode}%

\usepackage{amsmath,amssymb,bm,amsfonts}
\usepackage{fancyhdr,graphicx,pict2e}
\usepackage[greek,english]{babel}
\usepackage{natbib}
\usepackage{anysize}
\usepackage{setspace}
\usepackage[mathlines]{lineno}
\usepackage{amsthm}
\usepackage{tikz}
\usepackage{tikz-3dplot}
\usepackage{mathtools}
\usepackage{listings}
\usepackage{color} 
\definecolor{mygreen}{RGB}{28,172,0} 
\definecolor{mylilas}{RGB}{170,55,241}

\marginsize{2cm}{2cm}{2cm}{2cm}
\definecolor{lyell}{cmyk}{0,0,0.2,0}

\definecolor{dred}{rgb}{0.85,0,0}

\definecolor{dgreen}{rgb}{0,0.4,0}

\definecolor{dbrown}{cmyk}{0,0.30,0.45,0.64}

\definecolor{redy}{rgb}{0.7,0.6,0.6}

\definecolor{redb}{rgb}{0.7,0,0.7}

\definecolor{light}{gray}{.75}
\definecolor{dark}{gray}{.5}

\newcommand{\bc}{\begin{center}}
\newcommand{\ec}{\end{center}}
\newcommand{\bpic}{\begin{picture}}
\newcommand{\epic}{\end{picture}}
\newcommand{\ba}{\begin{array}}
\newcommand{\ea}{\end{array}}
\newcommand{\be}{\begin{eqnarray}}
\newcommand{\ee}{\end{eqnarray}}
\newcommand{\beq}{\begin{eqnarray*}}
\newcommand{\eeq}{\end{eqnarray*}}

\newcommand{\hsp}{\hspace*{\fill}}

\renewcommand{\d}{\mathrm{d}}

\newcommand{\e}{\mathrm{e}}

\newcommand{\lra}{\longrightarrow}

\newcommand{\dee}{\partial}
\newcommand{\del}{\boldsymbol{\nabla}}

\newcommand{\mb}[1]{\bm{#1}}

\newcommand{\R}[1]{\mathbb{R}^{#1}}

\renewcommand{\pi}{\textrm{\greektext p}}

\renewcommand{\theequation}{\arabic{equation}}

\begin{document}

\title[Diffusion around a sphere]{\center Nanoparticle uptake by a semi-permeable, spherical cell from an external planar diffusive field. II. Numerical study of temporal and spatial development validated using FEM.}

\author[1,2]{\fnm{Sandeep} \sur{Santhosh Kumar}}\email{ssanthoshkum@student.unimelb.edu.au}
\equalcont{These authors contributed equally to this work.}

\author*[1]{\fnm{Stanley J.} \sur{Miklavcic}}\email{stan.miklavcic@unisa.edu.au}
\equalcont{These authors contributed equally to this work.}

\affil*[1]{\orgdiv{Phenomics and Bioinformatics Research Centre, UniSA STEM}, \orgname{University of South Australia}, \orgaddress{\city{Adelaide}, \postcode{5095}, \state{SA}, \country{Australia}}}

\affil[2]{\orgdiv{School of Mathematics and Statistics}, \orgname{University of Melbourne}, \orgaddress{\city{Parkville}, \postcode{3010}, \state{VIC}, \country{Australia}}}
\abstract{In this paper we present a mathematical study of particle diffusion inside and outside a spherical biological cell that has been exposed on one side to a propagating planar diffusive front. The media inside and outside the spherical cell are differentiated by their respective diffusion constants. A closed form, large-time, asymptotic solution is derived by the combined means of Laplace transform, separation of variables and asymptotic series development. The solution process is assisted by means of an effective far-field boundary condition, which is instrumental in resolving the conflict of planar and spherical geometries. The focus of the paper is on a numerical comparison to determine the accuracy of the asymptotic solution relative to a fully numerical solution obtained using the finite element method. The asymptotic solution is shown to be highly effective in capturing the dynamic behaviour of the system, both internal and external to the cell, under a range of diffusive conditions.}


\maketitle

\date{\today}

\section{Introduction} \label{sec:introduction}

The phenomenon of diffusion has its classical origin in the 3D movement of particles in liquids or gases driven by a concentration gradient \cite{crank1956}. More recently, the principles underlying diffusion manifest themselves in other applied fields, such as sociology and economics, as well as in other circumstances in chemistry and physics. A practical example in biotechnology where diffusion is a key driving process is in molecular communication in which organic molecules become the carriers of information between cells~\cite{farsad2016}. Another example application may also be highlighted: with the advent of nanotechnology, engineered nanoparticles are used as drug delivery agents~\cite{allen2004}.

Accompanying these diverse applications is an increased number of complex, diffusive scenarios. The specific complexity of interest arises in cell-cell and particle-cell communication: a cell being exposed to the diffusing agents from one direction. In a recent review, Ashraf \textit{et al.}~\cite{ashraf2020} pointed out that different outcomes could result from different analysis methods for the same system. In the same vein, different conclusions are possible, about a cell's propensity to take up particles from its environment, depending on the nature of exposure. Circumstantial support for this can be gleaned from experimental studies featuring a directional dependence of exposure~\cite{cui2016}. The more frequently encountered situation, in nature and likely also in the laboratory, is arguably that of unidirectional exposure, which calls into question the relevance of many model calculations. Early studies \cite{crank1956,rashevsky1948,philip1964,mild1971}, adopted the simplifying assumption of spherical symmetry. This simplifying assumption, if used in modern experimental situations would inevitably lead to a misrepresentation of a cell's own involvement in particle uptake.

Recent modelling efforts to quantify (nano)particle uptake by biological cells that circumvent the spherical symmetry assumption have varied from spatially featureless, multiprocess kinetic models \cite{Sorrell2014,West2021}, to fully numerical, 3D simulation models \cite{Hinderliter2010,AlObaidi2015,Friedmann2016}. Sch\"{a}fer and collaborators \cite{Schafer2019,Schafer2020} expounded a so-called transfer function model, describing diffusion within a spherical biological cell subject to general time and position dependent boundary conditions on its semipermeable surface. The transfer function model was based on a Laplace transform-eigenfunction expansion method. Despite the generality of the method ensued by the arbitrary temporally- and spatially-dependent boundary condition applied to the cell's surface, the method falls short of the needs of the present problem. The specific case, addressed herein and in \cite{Miklavcic2024}, of a semi-permeable biological cell exposed to a field of diffusing particles generated by a planar source located in the unbounded, exterior domain, requires the concurrent solution of the exterior diffusion problem in order to satisfy the continuity conditions on the cell surface.

In \cite{Miklavcic2024} a closed-form, approximate solution of this combined diffusion problem was presented. The solution supported the hypothesis that the time dependence as nanoparticle uptake would explicitly reflect the unidirectional (as opposed to isotropic) nature of exposure. Moreover, it was shown that only through a scaling or modulation of this time dependence did the properties of the cell itself (its permeability, internal diffusion constant and its capacity to absorb particles) come into play. A comparison with a corresponding isotropic model, for the same cell's affinity, illustrated the importance of a proper consideration of the external condition in determining a cell's response to externally supplied substances.

Due to space restrictions, no assessment of the accuracy of the asymptotic solution was included in \cite{Miklavcic2024}. Also absent was a detailed study of either the internal or external time and space dependent diffusion process. We address these two deficiencies in this paper. It will be seen that the asymptotic expressions derived for the large-time particle concentration inside and around the cell are remarkably accurate when compared with a full numerical solution of the same system using the finite element method (FEM). The qualitative and quantitative agreement gives confidence to the subsequent application of the asymptotic solution to derived quantities of experimental relevance.

This report of our comparative study is organized as follows. In the next section we outline the physical problem under study and its representative mathematical model. The solution of the mathematical problem is summarized in Section \ref{sec:solution}, while full mathematical details are relegated to Appendices. The solution quoted here are fully equivalent to, but expressed more succinctly than, the explicit asymptotic expression given in~\cite{Miklavcic2024}. The alternative format here is arguably more advantageous for numerical purposes. Particulars about the FEM calculations are provided in Section \ref{sec:fem}. In Section \ref{sec:results} we compare full numerical (FEM) solutions of the problem with our numerically implemented asymptotic solution, under a range of parametric conditions. Both sets of results are shown as qualitative, 2D heat maps, while specific quantitative accuracy measures of the approximation are given in accompanying tables. The paper concludes in Section \ref{sec:conc} with some summary remarks and suggestions for future work.

\begin{figure}[t!]
    \centering
 \begin{tikzpicture}[scale=0.8,  rotate=90]
   \def\w{8}
   \draw[black,line width=1 mm,dashed] (5.15,-\w) -- (5.15,\w);
   \draw[black,line width=0.5mm,dashed] (4.25,-\w) -- (4.25,\w);
   \draw[black,line width=0.25mm,dashed] (3.5,-\w) -- (3.5,\w);
   \draw[black,line width=0.125mm,dashed] (2.75,-\w) -- (2.75,\w);
   \fill[draw=black,fill=black!10,line width=0.5mm,dashed] (0,0) circle (2cm);
    \draw[black, dotted, line width=0.25mm] (0,0) circle (4.75 cm);
   \fill[black!10] (5.2,\w) rectangle (8.5,-\w);
   \node[align=center] at (-0.35,0.5) {Cell~~~\\ (Region $II$)\\$c_{II},\,D_{II}$};
   \draw[-,draw=black,line width=0.35mm] (0,0) -- (5.18,0) node[at end, left, xshift=-0.1cm,yshift=0.25cm] {$z_0$} ;
   \draw[-,draw=black] (0,0) -- (1.4,-1.4) node[below right, pos=0.7] {$a$} ;
    \draw[->] (0,0) --(8,0) node[right] {$z$} ;
   \draw[->] (0,0) --(0,-7) node[right] {$x$} ;
    \draw[-,draw=black] (0,0) -- (4.385,-1.827) node[below right, pos=0.7, xshift=0.2cm,yshift=0.4cm] {$R$} ;
   \node[align=center] at (1.25,4.5) { External Medium \\ (Region $I$) \\ $c_{I},\,D_{I}$};
   \node[align=left] at (7,-3) {Particle\\~~~~ Reservoir\\~~~~~~~~~~$C_0$};
   \draw[|->,black, line width=0.5mm] (5.15,-7) -- (3.75,-7) node[right, pos=0.85, xshift=  0.25cm] {$\mathbf{N_{surf}}$};
   \def\a{2}
   \coordinate (circleCenter) at (-1,-1);
        \coordinate (pointOnCircle) at (-1.414,-1.414);
        \draw[|->,black, line width=0.5mm] (pointOnCircle) -- ($(pointOnCircle)!-\a!(circleCenter)$) node[above, pos=0.85, xshift=  0.25cm] {$\mathbf{N}$};
\end{tikzpicture}
   \caption{The physical system being modelled. A reservoir of particles at concentration $C_0$ lies in the infinite half-space $z \ge z_0$ and produces a diffusive front (indicated schematically by the horizontal dashed lines) directed toward down ($z<z_0$) in the direction $\mathbf{N}_{surf}=(0,0,-1)$, and encounters a spherical cell of radius $a \ll z_0$ centered at the origin of a Cartesian coordinate system, $(0,0,0)$. The outward pointing unit vector normal to the biological cell is defined as $\mathbf{N}=\widehat{\mathbf{r}}$.}
    \label{fig:system_cross section}
\end{figure}
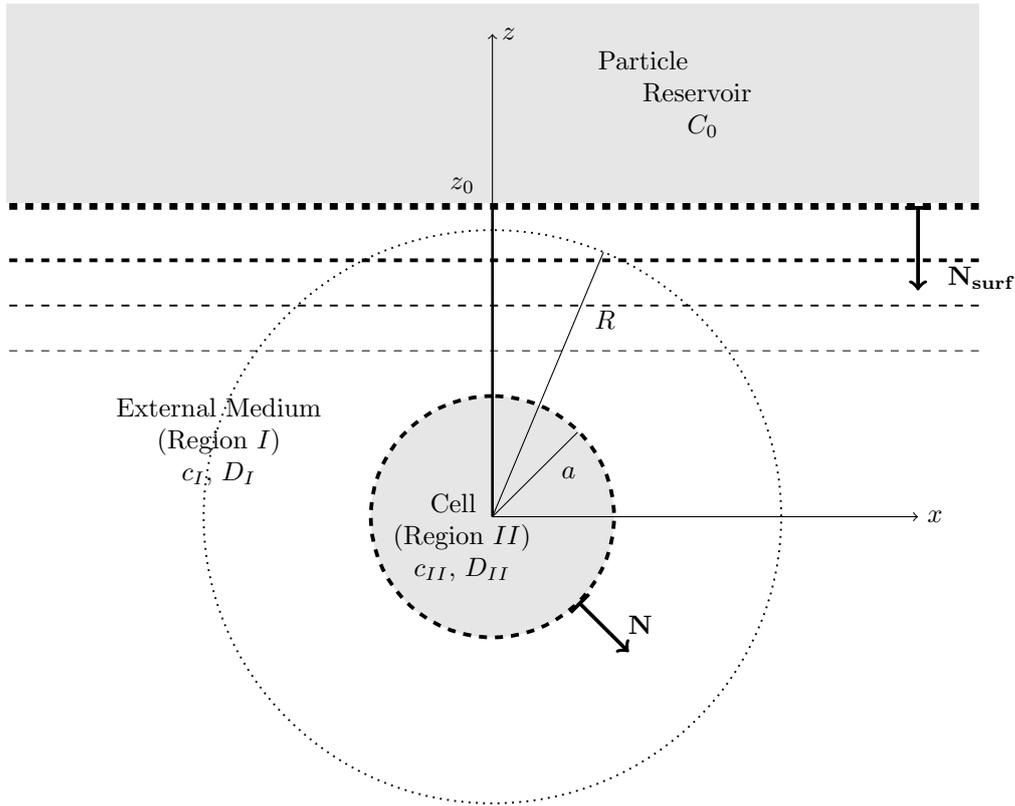

\section{Method} \label{sec:model}
\subsection{The Physical Problem}

Figure~\ref{fig:system_cross section} depicts the physical circumstances of the system we are modelling. We follow the variable notation introduced in \cite{Miklavcic2024}. A single biological cell (Region $II$) of spherical radius $a$ and internal diffusion constant $D_{II}$ is located in a liquid medium (Region $I$) characterized by a diffusion constant $D_{I}$. A distance $z_0 \gg a$ away from the biological cell centre is a planar boundary to an infinite half-space reservoir of particle solutes, held at a fixed number concentration of $C_0$. Initially, both the cell interior and the cell exterior, below the planar boundary of the reservoir, are free of particles. Over time the solutes diffuse into Region $I$, initially as a plane diffusive front which approaches the cell (Region $II$), as illustrated by the thin dotted lines in Figure~\ref{fig:system_cross section}. As the diffusive front approaches the biological cell the concentration contours (the level sets of particle concentration) bend toward the cell to eventually match the cell's curvature. The particles then diffuse into the cell and propagate internally either more slowly or faster than they would in the cell exterior, depending on the magnitude of the ratio $A = D_{I}/D_{II}$.

\subsection{The Mathematical Model}

For convenience, the biological cell is positioned at the origin ($\mb{r}=(x,y,z)=(0,0,0)$) of a 3D coordinate system. Adopting the usual coordinate structure, the interface of the particle reservoir is located at $z=z_0$, with the plane having the normal unit vector, $\mb{N}_{surf}=(0,0,-1)$, directed into Region $I$. We denote the local concentration of solute particles at time $t$ in the exterior medium, $z<z_0$, by $c_{I}(\mathbf{r},t)$ , while the concentration interior to the cell at time $t$ is $c_{II}(\mathbf{r},t)$.

Although more general cell properties were considered in \cite{Miklavcic2024}, for the purposes of this report we shall simply adopt the conditions of continuity of particle concentration and normal particle flux at the infinitesimally thin cell boundary, but otherwise assume arbitrary diffusion constants, $D_{I}$ and $D_{II}$. The governing equations in the two regions are thus,
\begin{eqnarray}
\dfrac{\partial c_{I}}{\partial t} &=& D_{I}\nabla^2c_{I} = D_{I}\left[\dfrac{1}{r^2}\dfrac{\dee}{\dee r}\left(r^2\dfrac{\dee c_{I}}{\dee r}\right)
+\dfrac{1}{r^2 \sin \phi}\dfrac{\dee}{\dee \phi}\left(\sin \phi \dfrac{\dee c_{I}}{\dee \phi}\right)
+\dfrac{1}{r^2 \sin^2 \phi} \dfrac{\dee^2 c_{I}}{\dee \theta^2}\right], \label{diffRegI} \vspace{0.85cm}\\
&& \hspace{6.5cm} r>a, \quad \phi\in[0,\pi], \quad \theta \in [0,2\pi], \quad t>0, \nonumber \\
 \nonumber \\
\dfrac{\partial c_{{II}}}{\partial t} &=& D_{II}\nabla^2c_{II} = D_{II}\left[\dfrac{1}{r^2}\dfrac{\dee}{\dee r}\left(r^2\dfrac{\dee c_{II}}{\dee r}\right)
+\dfrac{1}{r^2 \sin \phi}\dfrac{\dee}{\dee \phi}\left(\sin \phi \dfrac{\dee c_{II}}{\dee \phi}\right)
+\dfrac{1}{r^2 \sin^2 \phi} \dfrac{\dee^2 c_{II}}{\dee \theta^2}\right], \label{diffRegII} \vspace{0.85cm}\\
&& \hspace{6.5cm} r<a, \quad \phi\in[0,\pi], \quad \theta \in [0,2\pi], \quad t>0. \nonumber
\end{eqnarray}

These are complemented by initial and boundary conditions to ensure a unique solution. The initial condition specifies that both regions are devoid of particles at time $t=0$:
\begin{equation}
\left\{
\begin{array}{l}
c_{I}(\mb{r}, t = 0) = 0, \quad \quad \text{ \ in Region }I,\\
c_{II}(\mb{r}, t = 0) = 0, \quad \quad \text{ \ in Region }II.
\end{array}
\right. \label{icond}
\end{equation}
One boundary condition represents the continuity of the normal components of the solute surface fluxes across the cell boundary, $r=|\mb{r}|=a$:
\begin{equation}
\left[\mb{N} \cdot  \left(-D_{II} \del c_{II}\right) \right](r=a,\phi, t) = \left[\mb{N} \cdot \left(-D_{I}\del c_{I}\right) \right](r=a,\phi,t), \quad \phi\in[0,\pi],~t > 0,
\end{equation}
where $\mb{N}=\widehat{\mb{r}}$ is the outward pointing unit normal to the biological cell boundary $r=a$ (Figure \ref{fig:system_cross section}). The second boundary condition describes the condition of continuity of the concentration across the cell surface.:
\begin{equation}
c_{II}(r=a,\phi,t) = c_{I}(r=a,\phi,t), \quad \quad \phi\in[0,\pi],~t > 0. \hsp
\label{cond:conc}
\end{equation}
The final boundary conditions reflect, on the one hand, the continual provision of solutes at the interface with the particle reservoir, at $z=z_0$ and on the other hand, the fact that far from both the cell and the plane source, the concentration of solutes approaches zero:
\begin{multline}
\left\{
\begin{array}{l}
c_{I}(\mb{r},t) = C_0, \hspace{4.5cm} z=z_0, \quad \quad x,y \in \R{2}, \quad t>0,\\
c_{I}(\mb{r},t) \lra 0, \hspace{4.4cm} z \rightarrow -\infty, \quad x,y \in \R{2}, \quad t>0.
\end{array}
\right. \hsp \label{plnbc}
\end{multline}

\bigskip

\subsection{Non-dimensional System of Equations} \label{sec:nondim}
To simplify both the mathematical analysis and the numerical study to follow we introduce the following dimensionless variables and parameters,
\begin{equation}
\widetilde{\mathbf{r}}= \dfrac{\mathbf{r}}{a}; \quad \rho = \dfrac{r}{a}; \quad \tau = \dfrac{D_{II} t}{a^2}; \quad C_1= \dfrac{c_{I}}{C_0}; \quad C_2= \dfrac{c_{II}}{C_0}; \quad b = \dfrac{z_0}{a}; \quad A = \dfrac{D_I}{D_{II}}.
\end{equation}
Noting also that axisymmetry eliminates any dependence on the azimuthal angle, $\theta$, the governing equations reduce to
\begin{eqnarray}
\dfrac{\partial C_{1}}{\partial \tau} &=& \left[\dfrac{1}{\rho^2}\dfrac{\dee}{\dee \rho}\left(\rho^2\dfrac{\dee C_{1}}{\dee \rho}\right)
+\dfrac{1}{\rho^2 \sin \phi}\dfrac{\dee}{\dee \phi}\left(\sin \phi \dfrac{\dee C_{1}}{\dee \phi}\right) \right], \quad \quad \rho>1, \label{diffRegI} \vspace{0.85cm}\\
 \nonumber \\
\dfrac{\partial C_{2}}{\partial \tau} &=& A\left[\dfrac{1}{\rho^2}\dfrac{\dee}{\dee \rho}\left(\rho^2\dfrac{\dee C_{2}}{\dee \rho}\right)
+\dfrac{1}{\rho^2 \sin \phi}\dfrac{\dee}{\dee \phi}\left(\sin \phi \dfrac{\dee C_{2}}{\dee \phi}\right)\right], \quad 0<\rho<1, \label{diffRegII} 
\end{eqnarray}
for $\phi\in[0,\pi], \tau>0$. In addition, we state the non-dimensional versions of the continuity conditions at the cell surface,
\begin{multline}
\left\{
\begin{array}{l}
\dfrac{\partial C_{2}}{\partial \rho}(\rho=1,\phi, \tau) = A\dfrac{\partial C_{1}}{\partial \rho} (\rho=1,\phi,\tau), \\
C_{2}(\rho=1,\phi,\tau) = C_{1}(\rho=1,\phi,\tau), \hsp
\end{array}
\right.  \quad \quad \phi\in[0,\pi],~\tau > 0. \hsp \label{plnbc1}
\end{multline}
Finally, we have the far field conditions which, pointedly, are expressed in Cartesian geometry,
\begin{multline}
\left\{
\begin{array}{l}
C_{1}(\widetilde{\mb{r}},\tau) = 1, \hspace{4.7cm} \widetilde{z}=b,\\
C_{1}(\widetilde{\mb{r}},\tau) \lra 0, \hspace{4.4cm} \widetilde{z} \rightarrow -\infty,
\end{array}
\right.  \quad \quad (\widetilde{x},\widetilde{y}) \in \R{2}, \quad \tau>0, \hsp \label{plnbc2}
\end{multline}
where we have used the important dimensionless parameters, $A$ and $b$, introduced above.

\subsection{Re-specification of the Far-field Boundary Condition} \label{farfield}
In reducing the problem to one displaying axisymmetry (with no dependence on azimuthal angle), the first of Eqs \eqref{plnbc2} nevertheless introduces a conflict of spherical and planar geometries. We may resolve this conflict by noting that \emph{to a very good approximation} the biological cell will not have any influence on the concentration at points far from the cell. Hence, if the radius of the biological cell were very much smaller than the distance between the planar source and the cell centre, $a \ll z_0$, the time-dependent concentration of diffusing particles on the surface of a virtual (\emph{i.e.}, fictitious) sphere of radius $R$ satisfying $z_0 \ge R \gg a$, will be that determined by unidirectional diffusion alone. This is likely a better approximation when diffusion in the cell interior ($D_{II}$) is slower than diffusion in the external medium ($D_{I}$). In short, this suggests an alternative to boundary conditions Eqs \eqref{plnbc2}.

It is easily shown that, for particle propagation from the planar source in the direction of negative $z$ ($z \le z_0$), the solution of the dimensionless, unidirectional diffusion problem (with respect to independent variables $t$ and $z$),
\begin{multline}
\left\{
\begin{array}{ll}
\dfrac{\dee C_{1,\infty}}{\dee \tau} = A \dfrac{\dee^2 C_{1,\infty}}{\dee \widetilde{z}^2}, \hspace{4cm} &\text{for} \quad \widetilde{z} < b, \quad \tau > 0, \hsp\\
C_{1,\infty}(\widetilde{z},\tau) = 1, &\text{for} \quad \widetilde{z}=b, \quad \tau \ge 0, \hsp\\
C_{1,\infty}(\widetilde{z},\tau) \lra 0, &\text{as } \quad \widetilde{z} \lra -\infty, \quad \tau \ge 0, \hsp
\end{array}
\right. \label{OneDprob}
\hsp
\end{multline}
is~\cite{crank1956},
\begin{equation}
C_{1,\infty}(\widetilde{z},\tau) = \left(1 - \text{erf}\left(\dfrac{(b-\widetilde{z})}{\sqrt{4 A \tau}}\right)\right), \quad \quad \quad -\infty < \widetilde{z} \le b, \quad \tau \ge 0, \hsp \label{OneDsoln}
\end{equation}
where $\text{erf}(y) = \dfrac{2}{\sqrt{\pi}}\int_0^y \e^{-x^2} \d x$ is the error function~\cite{abram1965}. Consequently, the pair of boundary conditions in Eq. \eqref{plnbc2} may be replaced with a single, inhomogeneous boundary condition at the virtual sphere $\rho=X=R/a \gg 1$:
\begin{equation}
C_{1}(\rho=X,\phi,\tau) = C_{1,\infty}(X \cos \phi,\tau),  \quad \quad \quad \phi \in[0,\pi], \quad \tau > 0. \hsp \label{OuterBC}
\end{equation}

\section{Asymptotic and Numerical Solution of the Diffusion Problem} \label{sec:solution}
\subsection{Asymptotic approximation}
The re-defined temporal-spatial diffusion problem is amenable to a Laplace transform~/~separation-of-variables solution approach. With a Laplace transform with respect to the time variable defined as~\cite{mf1953},
\begin{equation}
\overline{C}_{\varsigma} (\rho,\phi:p) = \mathfrak{L}\left\{C_{\varsigma}\right\} = \int_0^{\infty} \e^{-p\tau} C_{\varsigma} (\rho,\phi,\tau) \d \tau, \quad \quad \varsigma = 1,2,
\label{lplc}
\end{equation}
the governing equations and boundary conditions become, for $\phi \in [0,\pi]$,
\begin{equation}
\left\{
\begin{array}{ll}
p \overline{C}_{1} = \left[\dfrac{1}{\rho^2}\dfrac{\dee}{\dee \rho}\left(\rho^2\dfrac{\dee \overline{C}_{1}}{\dee \rho}\right)
+\dfrac{1}{\rho^2 \sin \phi}\dfrac{\dee}{\dee \phi}\left(\sin \phi \dfrac{\dee \overline{C}_{1}}{\dee \phi}\right) \right], \hspace{2.05cm} &1 < \rho < X, \vspace{0.65cm}\\
\dfrac{p}{A} \overline{C}_{2} = \left[\dfrac{1}{\rho^2}\dfrac{\dee}{\dee \rho}\left(\rho^2\dfrac{\dee \overline{C}_{2}}{\dee \rho}\right)
+\dfrac{1}{\rho^2 \sin \phi}\dfrac{\dee}{\dee \phi}\left(\sin \phi \dfrac{\dee \overline{C}_{2}}{\dee \phi}\right) \right], &0 < \rho < 1, \vspace{0.5cm}\\
\overline{C}_{1}(X,\phi,p) = \overline{C}_{1,\infty}(X \cos \phi, p), \hsp \vspace{0.2cm} &\\
\dfrac{\dee \overline{C}_{2}}{\dee \rho}(1,\phi, p) = A\dfrac{\dee \overline{C}_{1}}{\dee \rho}(1,\phi,p), &\hsp \vspace{0.2cm}\\
\overline{C}_{2}(1,\phi,p) = \overline{C}_{1}(1,\phi,p). &\hsp
\end{array}
\right. \label{TrnsfPrblm}
\end{equation}
In Eq. \eqref{TrnsfPrblm}, $p$ is the Laplace parameter corresponding to dimensionless time $\tau$, and we have utilized the initial conditions, Eq. \eqref{icond} in the transform of the time derivative.

It may readily be confirmed that the separation-of-variables solution of Eq. \eqref{TrnsfPrblm} is given by
\begin{multline}
\left\{
\begin{array}{ll}
\overline{C}_{1}(\rho,\phi,p) = \sum_{n=0}^{\infty}\left[A_n i_n\left(\rho\sqrt{p/A}\right)+B_n k_n\left(\rho\sqrt{p/A}\right)\right]P_n(\cos \phi), &1 \le \rho \le X,\\
\overline{C}_{2}(\rho,\phi,p) = \sum_{n=0}^{\infty}D_n i_n\left(\rho\sqrt{p}\right)P_n(\cos \phi), \hspace{5.05cm} &0 \le \rho \le 1,\\
\end{array}
\right. \hsp \\ \hsp
\label{TrnsfSln}
\end{multline}
for $\phi \in [0,\pi]$.

In Eq. \eqref{TrnsfSln} $i_n$ and $k_n$ are the modified spherical Bessel functions of the first and second kind, of order $n$~\cite{abram1965,arfken2001},
\begin{equation}
i_n(z) = \sqrt{\dfrac{\pi}{2z}}I_{n+1/2}(z) \quad ; \quad \quad \quad k_n(z) = \sqrt{\dfrac{2}{\pi z}}K_{n+1/2}(z), \label{sphbssl}
\end{equation}
with $I_{n+1/2}$ and $K_{n+1/2}$ being the fractional order, modified Bessel functions of first and second kind. Furthermore, to obtain a finite solution defined on the closed interval $[0,\pi]$ we retain only $P_n(\cos \phi)$, the Legendre polynomials of the first kind \cite{abram1965}.

The expansion coefficients, $A_n$, $B_n$, and $D_n$, are chosen to satisfy the Laplace-transformed boundary conditions in Eq \eqref{TrnsfPrblm}. Upon substituting the separation-of-variables solution, Eq. \eqref{TrnsfSln}, into Eq. \eqref{TrnsfPrblm} and invoking orthogonality of Legendre functions we obtain a $3\times3$ linear algebraic system of equations which may be readily solved. The details of this process, while not complex, are lengthy and so are relegated to Appendices. Some explicit results are also given in \cite{Miklavcic2024}.

In order to return to the temporal domain, the solution in the Laplace transformed domain must be inverted. This is achieved through evaluating complex contour integrals of the Bromwich-type~\cite{mf1953},
\begin{equation}
C_{\varsigma}(\rho,\phi,\tau) = \mathfrak{L}^{-1}\left\{\overline{C}_{\varsigma}\right\} = \dfrac{1}{2 \pi i} \int_{\sigma - i \infty}^{\sigma + i \infty} \overline{C}_{\varsigma}(\rho,\phi,p) \e^{p \tau} \d p, \quad \quad \varsigma = 1,2. \label{LplcInv_Dfn}
\end{equation}
Inserting our separation-of-variables solution, $\overline{C}_{1}(\rho,\phi,p)$ and $\overline{C}_{2}(\rho,\phi,p)$, into Eq. \eqref{LplcInv_Dfn} we obtain, at least formally, the eigenfunction expansions,
\begin{eqnarray}
C_{1}(\rho, \phi, \tau) &=& \sum_{n=0}^{\infty} \mathfrak{L}^{-1}\left\{A_n i_n\left(\rho\sqrt{p/A}\right)+B_n k_n\left(\rho\sqrt{p/A}\right)\right\} P_n(\cos \phi), \quad 1 \le \rho \le X. \hsp \label{exterior}\\
C_{2}(\rho, \phi, \tau) &=& \sum_{n=0}^{\infty} \mathfrak{L}^{-1}\left\{D_n i_n\left(\rho\sqrt{p}\right)\right\} P_n(\cos \phi), \quad 0 \le \rho \le 1. \hsp \label{interior}
\end{eqnarray}

The above Laplace inversions are arguably impossible to achieve explicitly for all times $\tau$ since the Laplace parameter not only appears in the arguments of the explicit Bessel functions, it also appears deeply embedded in the coefficients $A_n$, $B_n$ and $D_n$. Consequently, as a compromise solution we seek only an asymptotic approximation valid for large times. The derivation of this asymptotic approximation is outlined in the Appendices (see also Section 3.2 of \cite{Miklavcic2024}). The final approximate solutions take the form of a double series valid for $0 \le \rho \le 1$ and a double series valid for $1 \le \rho \le X$. In each double series, the ``outer'' index denotes the order of the Legendre polynomial in the eigenfunction series expansion and hence represents the angular dependence, while the ``inner'' index denotes the order of the asymptotic approximation in time $\tau$ (for a given $n$). The explicit forms of these double series are exemplified by Eq. (27) in \cite{Miklavcic2024} for the time and space dependent particle concentration in the biological cell interior. For our purposes here, of numerical implementation, it suffices to simply represent the Legendre eigenfunction series in the forms
\begin{equation}\label{C_2 with g(x,m,tau)}
    C_1=\sum_{n=0}^{\infty}\sum_{k=0}^{\infty} \nu_k^{(n)}\,g(-X/\sqrt{A},k+n/2-1,\tau) P_n(\cos(\phi)),
\end{equation}
and
\begin{equation}\label{C_1 with g(x,m,tau)}
    C_2=\sum_{n=0}^{\infty}\sum_{k=0}^{\infty} \psi_k^{(n)}\,g(-X/\sqrt{A},k+n/2-1,\tau) P_n(\cos(\phi)),
\end{equation}
where the time-and space-dependent factors, $g(x,m,\tau)$, and the constants, $\nu_k^{(n)}$ and $\psi_k^{(n)}$, are derived in \ref{sec: Large time asymptotic approximation inside the cell} and \ref{sec: Large time asymptotic approximation  outside the cell}, and defined by Eqs~\eqref{eq: definition of G(x,m,tau)}, \eqref{eq: definition of nu_k}, and \eqref{eq: definition od psi_k}, respectively. The $g(x,m,\tau)$ functions involve confluent hypergeometric functions, which can be developed into asymptotic series in the time variable.

\subsection{Numerical implementation of asymptotic solution} \label{sec:numasymp}
Although they may be developed to arbitrary order, for the practical purpose of numerical implementation we truncate the double series given in Eqs~\eqref{C_2 with g(x,m,tau)} and \eqref{C_1 with g(x,m,tau)}. After some exploration we found that a good balance of computational accuracy and computational efficiency could be achieved (see further comments below) by considering terms up to and including $k=2$ and $n=5$, for the inner and outer summations, respectively. Hence, the finite double summation approximations, which are formally only valid for large times, that we've implemented in our comparison with the fully numerical finite element solution, are
\begin{equation}
C_1 \approx\sum_{n=0}^{5}\sum_{k=0}^{2} \nu_k^{(n)}\,g(-X/\sqrt{A},k+n/2-1,\tau) P_n(\cos(\phi)),  \quad 1 \le \rho \le X,   \quad 0 \le \phi \le \pi,
     \label{eq: Asym approx for plots C_2}
\end{equation}
for outside the cell,
and
\begin{equation}
C_2\approx\sum_{n=0}^{5}\sum_{k=0}^{2} \psi_k^{(n)}\,g(-X/\sqrt{A},k+n/2-1,\tau) P_n(\cos(\phi)),  \quad 0 \le \rho \le 1,   \quad 0 \le \phi \le \pi, \label{eq: Asym approx for plots C_1}
\end{equation}
for inside the cell.

\begin{figure}[t!]
    \centering
    \includegraphics[scale=1]{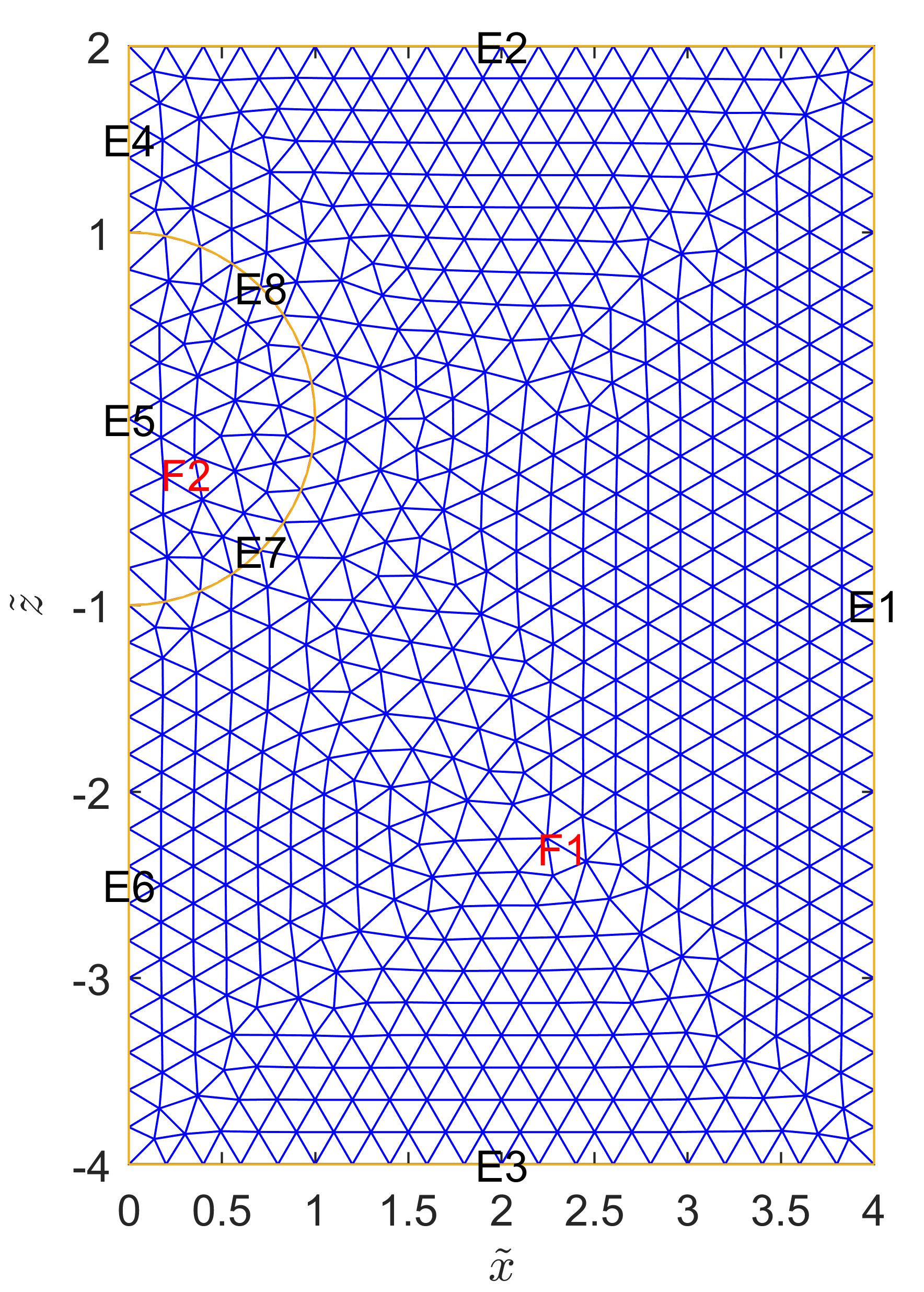}
    \caption{An illustration of the finite element method discretization mesh, here with $b=2$, $Q_1=6$ and $Q_2=4$. In our practical implementation we have generally adopted the dimensions $Q_1=50$ and $Q_2=20$, and with $b=X= 10$. The biological cell periphery is indicated by the yellow half-circle, and the cell exterior and interior are denoted F1 and F2, respectively.}
\begin{picture}(0,0)
\put(-175,250){$\left.\dfrac{\partial C_{1,2}}{\partial \Tilde{x}}\right|_{\Tilde{x}=0}=0$}
\put(-108,251){\linethickness{0.5pt}\vector(2.2,-7){24}}
\put(-108,251){\linethickness{0.5pt}\vector(1.3,-1){24}}
\put(-108,251){\linethickness{0.5pt}\vector(1.2,3.1){25}}
\put(30,340){$\left.C_1\right|_{\Tilde{z}=b}=1$}
\put(130,250){$\left.\dfrac{\partial C_1}{\partial \Tilde{x}}\right|_{\Tilde{x}=Q_2}=0$}
\put(130,150){$\left.\dfrac{\partial C_1}{\partial \Tilde{z}}\right|_{\Tilde{z}=(b-Q_1)}=0$}
\put(130,150){\linethickness{0.5pt}\vector(-5,-5){73}}
\put(130,250){\linethickness{0.5pt}\vector(-1,-1){36}}
\end{picture}
    \label{fig: FEM Mesh}
\end{figure}
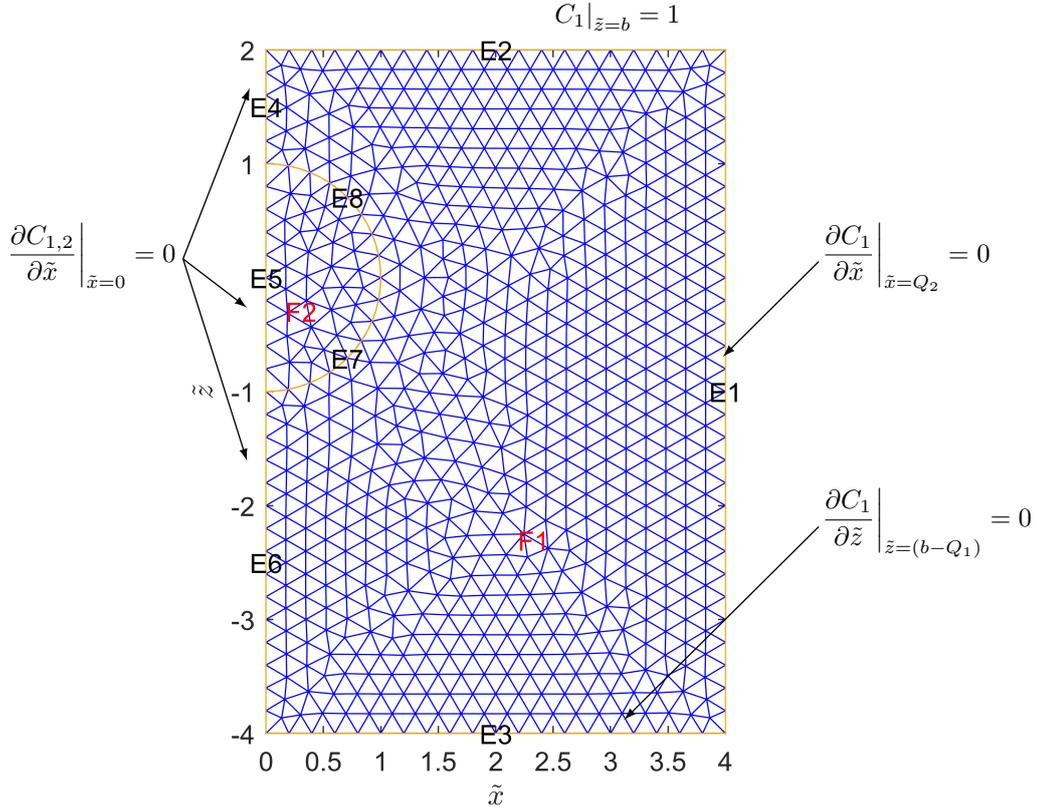
For the explicit cases we examined it was found that terms of higher order than $n=5$ in the Legendre function expansions contributed negligibly to the approximation.  The terms corresponding to $n=5$ for $k=0,1,2$ have a maximum magnitude of $10^{-4}$ for those values of $\tau$ examined in the case of $A>1$, and a maximum magnitude of $0.1$ for the corresponding case of $A<1$. The first three terms of $\psi_k^{(n)}$ and $\nu_k^{(n)}$ each contribute significantly to the approximations. If we omit the $\psi_2^{(n)}$ and $\nu_2^{(n)}$ terms, the resulting approximations are less accurate for a given $\tau$, but as $\tau$ increases the solutions nevertheless converge to the same degree of approximation one finds with the $k=2$ terms included. The process of explicitly computing the first three terms of $\psi_k^{(n)}$ and $\nu_k^{(n)}$ is explained in Appendix \ref{sec: Asymptotic analysis}. Naturally, the convergence behaviour and accuracy of the approximate solutions will depend on the specific values of $X$, $A$ and $\rho$ that are assigned.

In the numerical comparison below we can only consider the two regions relevant to the asymptotic approximations, which are summarized in Appendices \ref{sec: Large time asymptotic approximation inside the cell}~\&~\ref{sec: Large time asymptotic approximation  outside the cell}. That is, the numerical comparison pertains only to the region defined by the inequality $0< \rho < X$.

In the figures appearing in the next section we highlight contour lines which are curves joining points of equal concentration. These are the iso-contours. With the help of these one can readily appreciate (visualize) the time-dependent progress of the diffusive front.

\subsection{Finite element solution} \label{sec:fem}
As a benchmark solution with which to compare our asymptotic solution, we used the ``Partial Differential Equation Toolbox'' in \textsc{Matlab} to solve the system described in Section~\ref{sec:model}. The \verb|transient-axisymmetric| analysis module is used to exploit the azimuthal symmetry of the system around the $z$-axis. Moreover, it is sufficient to apply the finite element method to one half of the system cross-section, here taken to be the $xz$-plane. The spatial variables for this non-dimensional cross-section are defined as in Section \ref{sec:nondim},
\begin{equation}
     \Tilde{x}=\frac{x}{a}, \qquad \Tilde{z}=\frac{z}{a}.
\end{equation}

Figure~\ref{fig: FEM Mesh} shows a simple example of the elements generated by the PDE toolbox in \textsc{Matlab}. The element edges labeled \verb|E7| and  \verb|E8|, between $\Tilde{z}=-1$ and $\Tilde{z}=+1$, represent the boundary of our biological cell. We maintain a constant concentration condition, $C_1=1$, along the boundary edge \verb|E2|, which denotes the planar boundary generating the diffusive front. Along edges \verb|E1| and \verb|E3| we impose the condition of zero normal flux, $\partial C_1/ \partial \Tilde{x} = 0$ and $\partial C_1/ \partial \Tilde{z} = 0$, respectively. The latter boundary conditions are required since the PDE toolbox is incapable of treating infinite domain systems. Along the central, axis of symmetry we also impose a zero gradient condition, $\partial C_{1,2}/ \partial \Tilde{x} = 0$. These boundary conditions and their locations are shown in Figure~\ref{fig: FEM Mesh}. To best model the present infinite system, we capture as large an external region as possible.  We define parameters $Q_1$ (the length of \verb|E1|) and $Q_2$ (the length of \verb|E3|) as being the length and breadth, respectively, of a subset of the actual infinite spatial domain. Relevant edges are labeled \verb|E1| to \verb|E8|, and the exterior and interior regions are denoted \verb|F1| and \verb|F2|, respectively.

The FEM implementation involves a parameter $h_m$ which is the maximum distance between nodes; the smaller is $h_m$, the greater is the resolution.

\begin{figure}[t!]
  \centering
  \includegraphics[width=0.3\columnwidth]{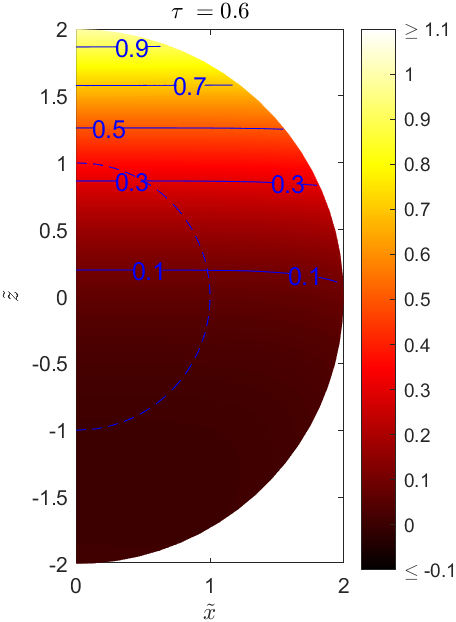}
  \hspace{0.15cm}
  \includegraphics[width=0.231\columnwidth]{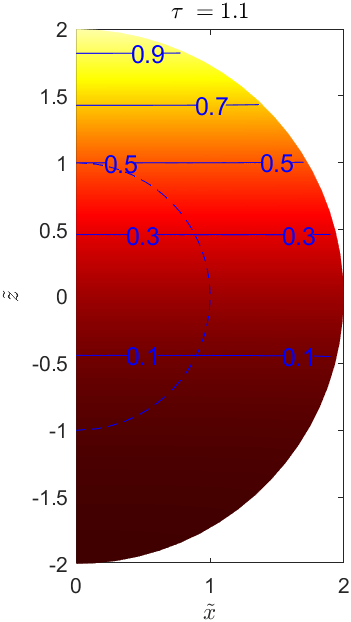}
  \hspace{0.15cm}
  \includegraphics[width=0.3\columnwidth]{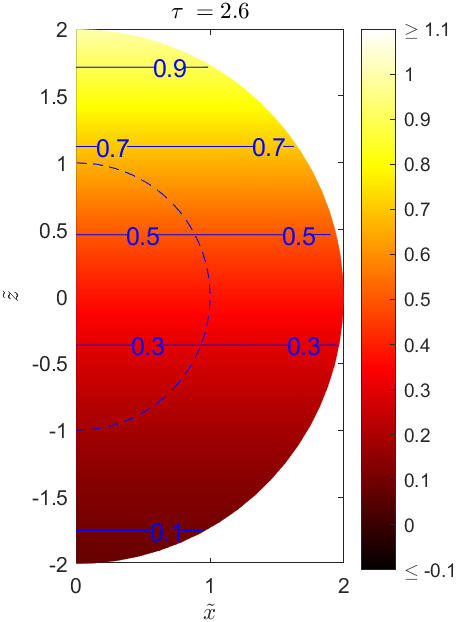}
\caption{Non-dimensional particle distributions inside and outside the cell predicted by the asymptotic approximation for $A=1$, and $b=X=2$. }
\label{fig:figure2}
 \end{figure}

\section{Results and Discussion} \label{sec:results}

\subsection{Preliminary remarks}
Before beginning a detailed numerical study, it is useful to establish the qualitative bounds within which we anticipate the asymptotic approximation to be valid. These qualitative confines of the model help to give a broader perspective to the specific quantitative comparison to follow. With the adopted model assumptions, and subsequent approximations made in the solution process, we anticipate the following limitations on accuracy.

Consider first the influence of the virtual sphere boundary. For all times \emph{if} $X$ is chosen such that $X \gtrsim O(1)$ then the asymptotic solution will cease to have meaning since the biological cell's presence will invariably affect the particle concentration at the position of the virtual sphere's boundary at $\rho=X$. The concentration distribution around the latter boundary will then be different at all times from those values assumed by the 1D planar solution, upon which the far field boundary condition is based. However, at the same time it has to be acknowledged that since a FEM simulation is necessarily also confined to a finite domain and hence is itself burdened by effective/approximate outer boundary conditions, it may be difficult to conclude from a comparison alone where and when, or even \emph{if}, a deficiency exists on the part of the asymptotic solution. To avoid or at least minimize any ambiguity, the majority of the results presented below were obtained assuming a radii ratio of $X = 10$. Nevertheless, for completeness, we also consider a few cases for which the ratio $X$ is as little as $2$.

A second consideration is that of time. Since the mathematical exercise produced a large-time, asymptotic solution we should anticipate that numerical inaccuracies will emerge in a numerical implementation for short and potentially medium times, revealed in a comparable FEM calculation. In other words, the numerical implementation will in general likely be in error for non-dimensional times $\tau \ll 1$ and potentially even for $\tau \approx O(1)$. It is in this second consideration, other factors being equal, that the FEM solution will be useful in revealing any errors. Since this latter prospect is an important focus of this paper, we shall compare the two numerical works for a range of non-dimensional times.

Given the first of the aforementioned two qualitative points we may ensure greatest numerical accuracy in the asymptotic approximation by setting $X=b$, which also conveniently reduces the extent of parameter space we are to explore to that of two dimensions. For the most part we adopt this choice; the only exception being our study of the effect of reducing the size of the virtual sphere boundary. Our investigation comprises a semi-quantitative, side-by-side comparison of temporal snapshots of particle distributions (in the form of heat maps) produced by the two methods. On a more quantitative level, for each graphical comparison we include an accompanying table of $L_{\infty}$, $L_1$ and $L_2$ measures \footnote{Relative as opposed to absolute measures could also be considered. However, as particle concentrations were of order 1 in the cases quantified (except at very short times) no greater appreciation of accuracy is achieved by including both sets of measures.}:
\begin{equation}
\left\{
\begin{array}{l}
L_{\infty} = \sup\left\{\left|C_{asym}-C_{\text{\tiny FEM}}\right|\right\},\\
L_{1} = \dfrac{1}{m}\sum_m \left|C_{asym}-C_{\text{\tiny FEM}}\right|, \label{measures}\\
L_{2} = \sqrt{\dfrac{1}{m}\sum_m \left|C_{asym}-C_{\text{\tiny FEM}}\right|^2},
\end{array}
\right.
\end{equation}
where $C_{\text{\tiny FEM}}$ and $C_{asym}$ denote the (dimensionless) concentrations predicted by the FEM and asymptotic solution methods, respectively. The indicated supremum and summations in Eq \eqref{measures} extend either over the biological cell interior, or the biological cell exterior (but within the virtual sphere of radius $\rho=X$). The number of discrete points sampled inside the biological cell was set at $m=50$, while external to the cell and within the virtual sphere $\rho=X$, $m=4472$ points were sampled unless otherwise indicated. Invariably, the non-dimensional spatial domain simulated in the FEM calculations was a rectangle of size $Q_1 \times Q_2 = 50 \times 20$ containing one half of the total, mirror-symmetric region, with a non-dimensional grid dimension (\textit{i.e.}, the maximum distance between FEM nodes) of $h_m=0.2$.

\begin{figure}[t!]
  \centering
  \includegraphics[width=0.3\columnwidth]{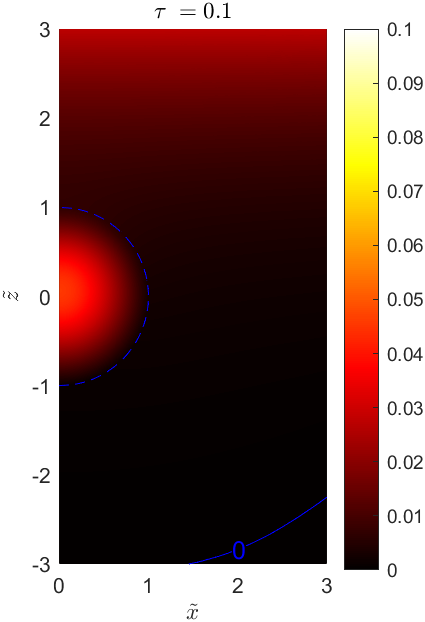}
  \hspace{0.15cm}
  \includegraphics[width=0.3\columnwidth]{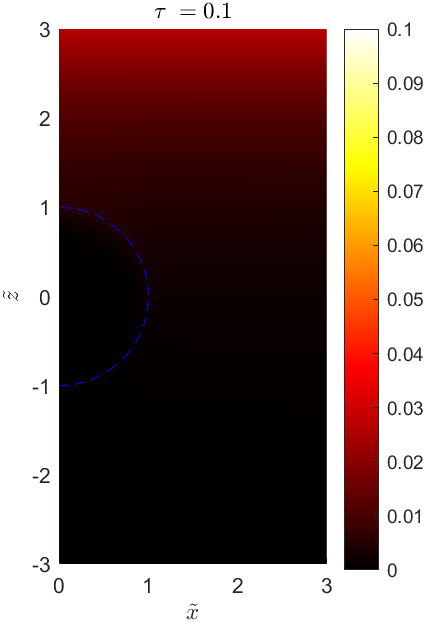}
 \caption{Comparison of non-dimensional particle distributions inside and outside the biological cell predicted by the asymptotic approximation (left) and FEM simulation (right) for $A=50$, and $b=X=10$, at a non-dimensional time of $\tau = 0.1$. The FEM simulation region is a rectangle of dimensions $Q_1 \times Q_2 = 50 \times 20$ with a grid spacing of $h_{m}=0.2$.}
  \label{fig:figure3}
 \end{figure}

\subsection{Numerical comparison}
As a convenient first test case, capitalizing on the assumed continuity of both concentration and normal flux across the membrane, we set the diffusion ratio to unity $A=1$ so that the rates of diffusion inside and outside the biological cell are equal. Under these conditions we would expect the cell, in principle, to be effectively transparent to the planar front. Consequently, we would expect the two predictions of particle concentration (FEM and asymptotic) to be consistent with the trivial case of the biological cell being completely absent. Thus, we would expect the particle iso-contours (contours joining points of equal concentration) to be parallel to each other and moreover parallel to the plane source of particles that produced them. These iso-contours would hence travel in the direction perpendicular to the planar source. These expectations are indeed realized as can be seen from the contours shown in Figure \ref{fig:figure2} for the particle distributions at three non-dimensional times. Incidentally, as a further test of the model's resilience we have here set the virtual sphere boundary to $X=b=2$. The FEM solution, which is not shown, is graphically indistinguishable from the asymptotic results shown in Figure \ref{fig:figure2}.

\begin{figure}[t!]
  \centering
  \includegraphics[width=0.3\columnwidth]{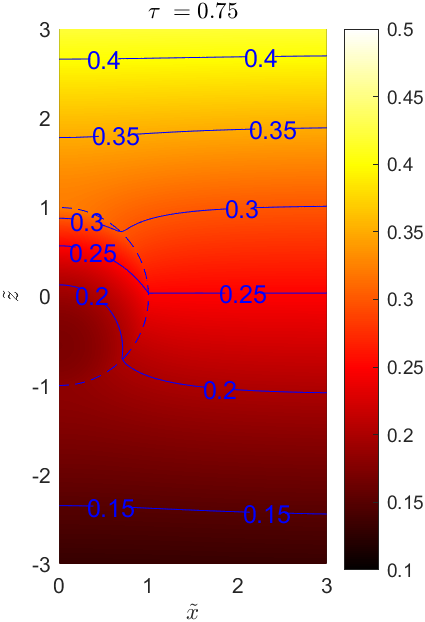}
  \hspace{0.15cm}
  \includegraphics[width=0.3\columnwidth]{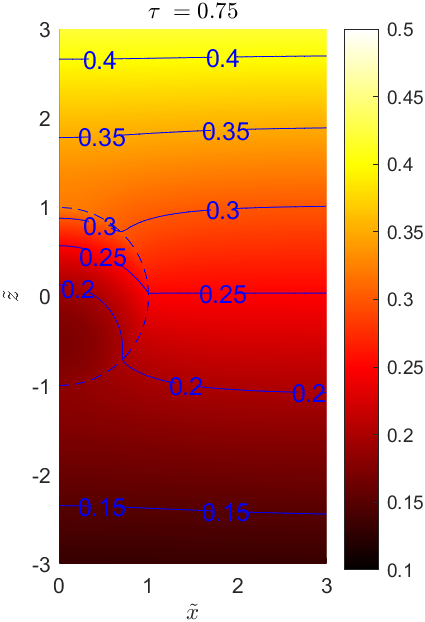} \\ \vspace{2mm}
  \includegraphics[width=0.3\columnwidth]{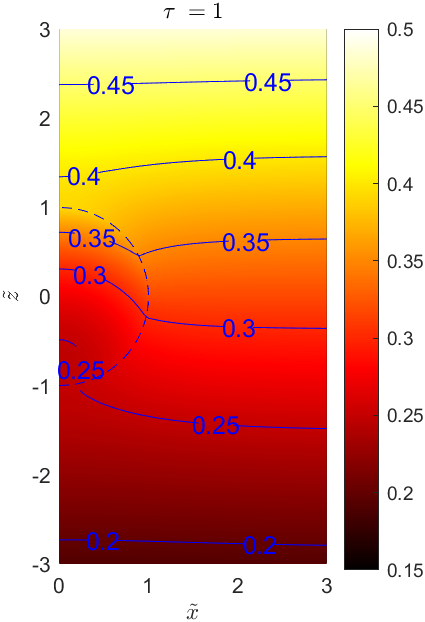}
  \hspace{0.15cm}
  \includegraphics[width=0.3\columnwidth]{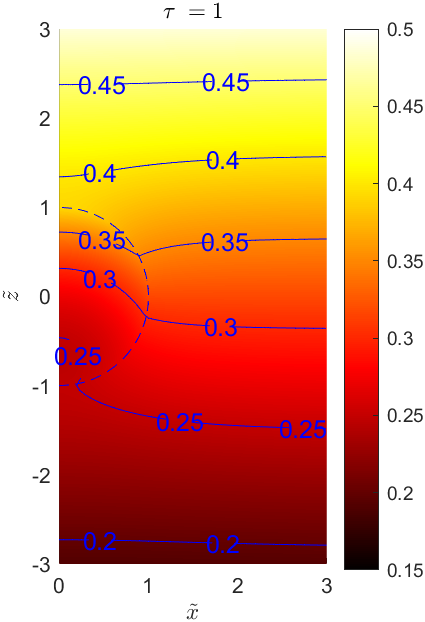} \\ \vspace{2mm}
  \includegraphics[width=0.3\columnwidth]{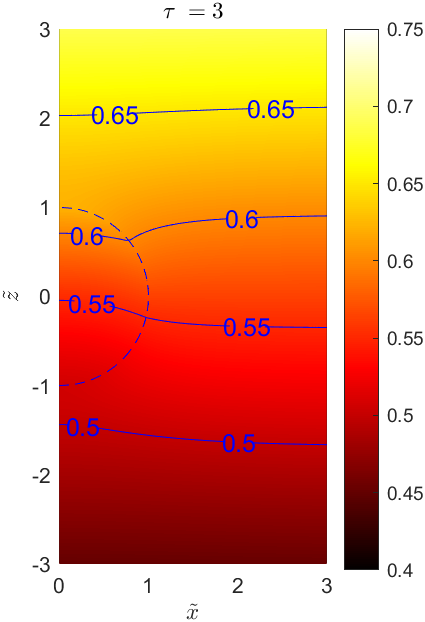}
  \hspace{0.15cm}
  \includegraphics[width=0.3\columnwidth]{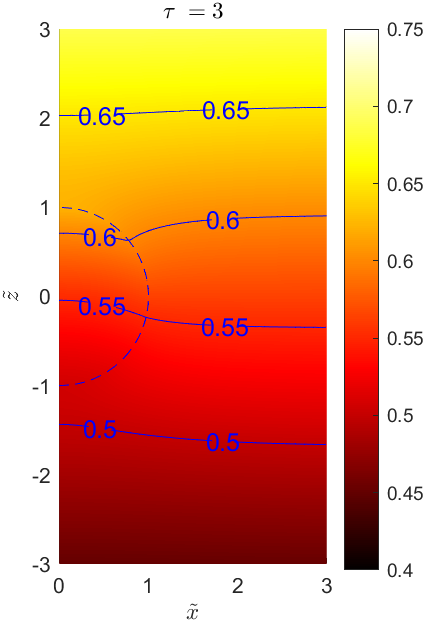} \\
 \caption{Comparison of non-dimensional particle distributions inside and outside the cell predicted by the asymptotic approximation (left) and FEM simulation (right) for $A=50$, $b=X=10$, and, top to bottom, at non-dimensional times of $\tau = 0.75, 1, \text{ and } 3$. The FEM simulation cell is a rectangle of dimensions $Q_1 \times Q_2 = 50 \times 20$ with a grid spacing of $h_{m}=0.2$.}
  \label{fig:figure4}
\end{figure}

For a diffusion constant ratio of $A=50$ (diffusion inside the biological cell is 50 times slower than diffusion outside the cell), we show in Figures \ref{fig:figure3}-\ref{fig:figure4} a sequence of snap-shots of these distributions from short to large dimensionless times. Confirming the second of our two qualitative expectations, we see that at $\tau = 0.1$ (Figure \ref{fig:figure3}) there is an obvious discrepancy between the two results, with the asymptotic theory falsely predicting an accumulation of particles inside the cell, despite the absence of particles in the cell exterior. However, by $\tau=0.25$ (data not shown), the prediction is already qualitatively in line with the FEM results, both inside and outside the biological cell. The agreement naturally improves as $\tau$ increases (Figure \ref{fig:figure4}): from $\tau = 0.75$ to $\tau=3$ in which interval the asymptotic solution is all but identical to the FEM solution. The improved agreement is confirmed quantitatively in Table \ref{tab:table1} which shows a $L_{\infty}$ error of $1.9\times10^{-4}$ inside the cell and a $L_{\infty}$ error of $1.6\times 10^{-4}$ outside the cell at $\tau=3$. Both $L_1$ and $L_2$ error measures are comparable (to each other) and suggest, when compared with the $L_{\infty}$ measure, that the maximum errors highlighted by $L_{\infty}$ are highly localized.

\begin{table}
\begin{tabular}{c|c|c|c|c|c|c}
  \hline
  $\tau$ & 0.1 & 0.25 & 0.5 & 0.75 & 1 & 3 \\ \hline
  $L_1(\rho < 1)$ & 1.79$\times10^{-2}$ & 6.25$\times10^{-3}$ & 2.73$\times10^{-3}$ & 8.51$\times10^{-5}$ & 3.18$\times10^{-4}$ & 7.60$\times10^{-5}$ \\
  $L_2(\rho < 1)$ & 2.32$\times10^{-2}$ & 8.20$\times10^{-3}$ & 3.54$\times10^{-3}$ & 1.24$\times10^{-4}$ & 4.06$\times10^{-4}$ & 8.30$\times10^{-5}$ \\
  $L_{\infty}(\rho < 1)$ & 4.50$\times10^{-2}$ & 1.61$\times10^{-2}$ & 6.87$\times10^{-3}$ & 3.49$\times10^{-4}$ & 9.30$\times10^{-4}$ & 1.93$\times10^{-4}$ \\ \hline
  $L_1(1 < \rho < X)$ & 2.44$\times10^{-2}$ & 4.14$\times10^{-3}$ & 7.09$\times10^{-4}$ & 2.67$\times10^{-4}$ & 1.56$\times10^{-4}$ & 6.66$\times10^{-5}$ \\
  $L_2(1 < \rho < X)$ & 5.13$\times10^{-2}$ & 6.99$\times10^{-3}$ & 1.18$\times10^{-3}$ & 4.81$\times10^{-4}$ & 2.61$\times10^{-4}$ & 8.23$\times10^{-5}$ \\
  $L_{\infty}(1 < \rho < X)$ & 2.65$\times10^{-1}$ & 3.32$\times10^{-2}$ & 5.44$\times10^{-3}$ & 2.38$\times10^{-3}$ & 1.15$\times10^{-3}$ & 1.63$\times10^{-4}$ \\
  \hline
\end{tabular}    \caption{Mean error measures ($L_1$, $L_2$, and $L_{\infty}$) as a function of non-dimensional time $\tau$, inside and outside the cell. The calculations were based on the case $A=50$, $b=X=10$, and for FEM grid dimensions of $Q_1 \times Q_2 = 50 \times 20$ and a grid spacing of $h=0.2$.}
    \label{tab:table1}
\end{table}

The slower rate of diffusion inside the biological cell compared with diffusion outside ($A=50 \Rightarrow D_{II} = D_{I}/50$) leads to an initial build-up in the particle concentration in the anterior half of the cell ($\phi \approx 0$) (this is certainly the case inside, but also outside, although to a lesser extent) so those profiles or iso-contours appear to lead those outside (\textit{i.e.}, are further along in $\Tilde{z}$ (see, for example, the $C=0.3$ iso-contour at $\tau=0.75$ or the $C=0.4$ iso-contour at $\tau=1$ in Figure \ref{fig:figure4}). However, in the posterior half of the cell ($\phi \approx \pi$), both inside and outside, the roles are reversed, with the external iso-contours leading (in $\Tilde{z}$) the corresponding internal iso-contours. This is evidenced by the $C=0.2$ concentration iso-contour at $\tau=0.75$ or the $C=0.25$ iso-contour at $\tau=1$ in Figure \ref{fig:figure4}. The external concentration gradient downstream and to the side of the cell, possesses a significant vector component in the direction toward the axis of symmetry, driving particles toward the cell's shadow region. With time and distance (downstream of the cell) the profile cross-sections tend once again to be linear, corresponding to the asymptotic, unidirectional planar front (see the $C=0.2$ contour at $\tau = 1$, middle pair of panels in Figure \ref{fig:figure4}). These behaviorial details are captured by both the asymptotic approximation and the FEM simulation. It is noteworthy that the tendency for the iso-contours to regain the planar form is consistent with and validates the use of the effective virtual sphere boundary condition at $X$, which takes on the value of the unidirectional solution, Eq \eqref{OneDsoln}, for all $\phi \in [0,\pi]$.

\begin{figure}[t!]
  \centering
  \includegraphics[width=0.3\columnwidth]{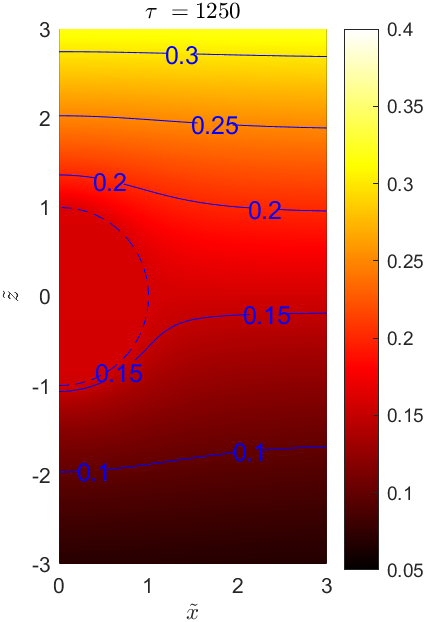}
  \hspace{0.15cm}
  \includegraphics[width=0.3\columnwidth]{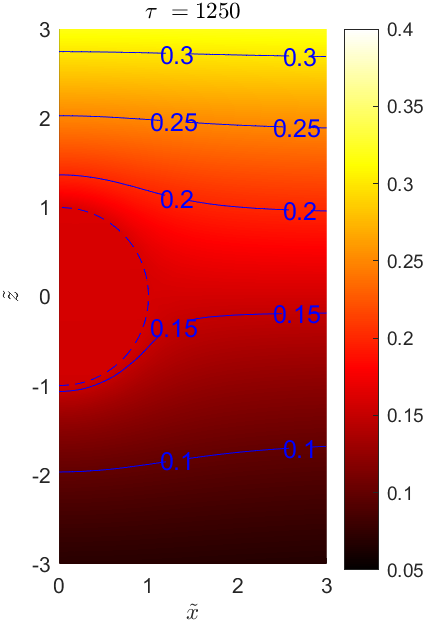} \\ \vspace{2mm}
  \includegraphics[width=0.3\columnwidth]{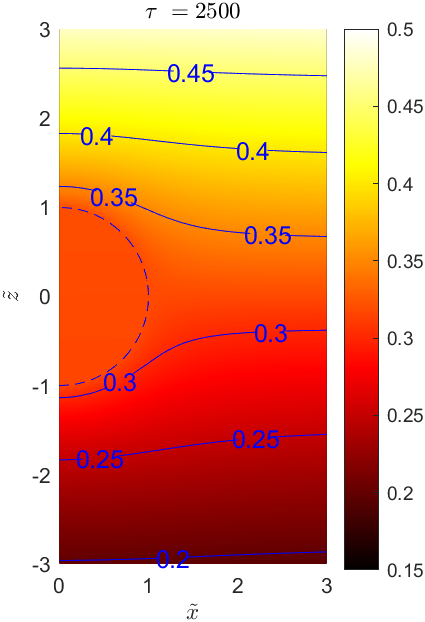}
  \hspace{0.15cm}
  \includegraphics[width=0.3\columnwidth]{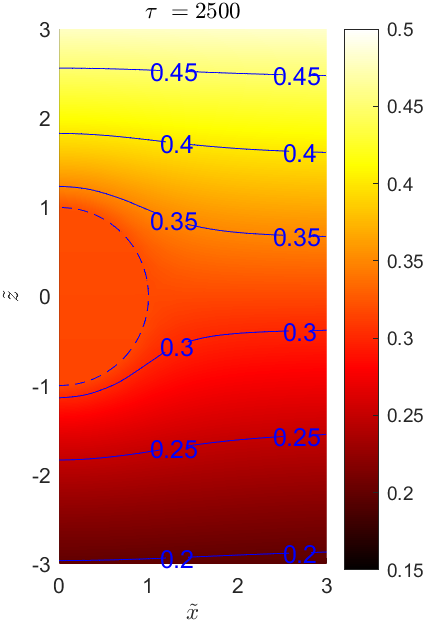} \\ \vspace{2mm}
  \includegraphics[width=0.3\columnwidth]{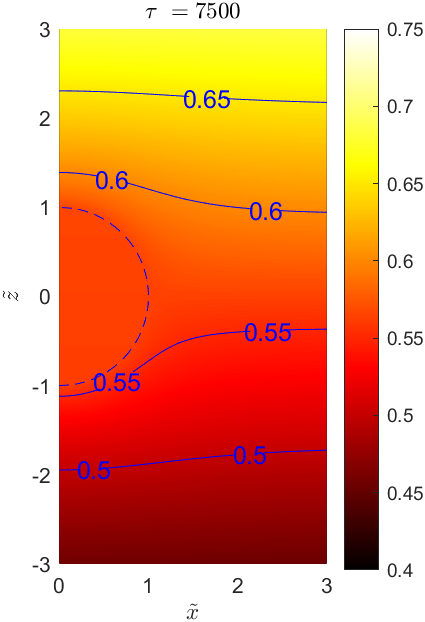}
  \hspace{0.15cm}
  \includegraphics[width=0.3\columnwidth]{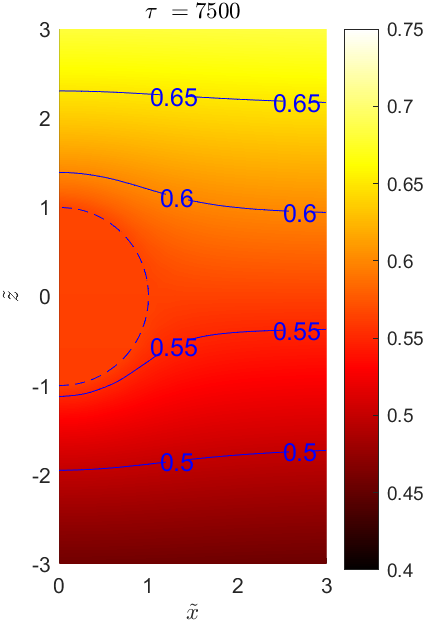} \\
 \caption{Comparison of non-dimensional particle distributions inside and outside the cell predicted by the asymptotic approximation (left) and FEM simulation (right) for $A=1/50$, $b=X=10$, and, top to bottom, at non-dimensional times of $\tau = 1250, 2500, \text{ and } 7500$. The FEM simulation cell is a rectangle of dimensions $Q_1 \times Q_2 = 50 \times 20$ with a grid spacing of $h_{m}=0.2$.}
  \label{fig:figure5}
\end{figure}

For the case $A=1/50$ ($D_{II} = 50 D_{I}$) diffusion inside the biological cell is faster than diffusion outside. Consequently, the relationship between iso-contour profiles inside and outside is somewhat the reverse of the case found with $A=50$, but is not a mirror reversal. Nevertheless, as we see from Figure \ref{fig:figure5}, the more rapid particle movement inside the cell results in a depletion zone in the cell anterior region ($\phi \approx 0$) and a build-up of particle concentration in the posterior region ($\phi \approx \pi$), by those particles that have been more rapidly transported through the cell. This more rapid internal movement is responsible for an accumulation of particles in the cell's shadow region. This higher level of accumulation generates a concentration gradient with a significant vector component directed away from the symmetry axis, which of course tends to equilibrate concentrations everywhere; the iso-contours some cell radii further downstream, again tend to be planar. However, this occurs at a slower rate overall compared with the $A=50$ example. In the cases shown, the diffusive propagation inside the cell is so rapid that no contours of the representative concentration increments are included.

In this case of internal diffusion being 50 times faster than external diffusion, agreement between the asymptotic theory and the FEM simulation still occurs but at significantly larger dimensionless times. The non-dimensional times required to obtain the same level of agreement as in the preceding case of $A=50$ are here two to four orders of magnitude greater that unity (compare the times of the entries with roughly equivalent accuracies ($L_1$ and $L_2$ measures) in Table \ref{tab:table1} and Table \ref{tab:table2}).

The significant difference in time orders is somewhat deceptive. In our numerical simulation we had followed the lead of Philip \cite{philip1964} and  Mild \cite{mild1971} and defined a dimensionless time based on the diffusion constant of the cell's interior, $\tau = D_{II} t/a^2$. Consequently, the orders of magnitude difference in non-dimensional time may be explained by noting that for the same physical time, $t$, other things being equal, the ratio of the dimensionless times in the two cases studied give
\begin{equation*}
\dfrac{\tau_{A>1}}{\tau_{A<1}} = \dfrac{\dfrac{0.02 D_1 t}{a^2}}{\dfrac{50 D_1 t}{a^2}} = \dfrac{1}{2500}.
\end{equation*}
Hence, in terms of non-dimensional times, scaled as we have done, we would expect a three order of magnitude large $\tau$ disparity for the same $t$ value. Equivalently stated, from Tables \ref{tab:table1} and \ref{tab:table2} we see, for example, almost identical accuracy was achieved in the outer region at $\tau=0.25$ in the case of $A=50$, as at $\tau = 625$ in the case of $A=1/50$. These levels of accuracy occurred approximately at
\begin{equation*}
t_{A>1} = \dfrac{\tau_{A>1} a^2}{D_{I}/A} \approx \dfrac{0.25 a^2}{D_{I}/50} = 12.5 \dfrac{a^2}{D_{I}},
\end{equation*}
in the case of $A=50$, and
\begin{equation*}
t_{A<1} = \dfrac{\tau_{A<1} a^2}{D_{I}/A} \approx \dfrac{625 a^2}{50 D_{I}} = 12.5 \dfrac{a^2}{D_{I}},
\end{equation*}
in the case of $A=1/50$. That is, the same level of agreement occurring at widely different non-dimensional times actually occur at the same real time.

\begin{table}
\begin{tabular}{c|c|c|c|c|c|c}
  \hline
  $\tau$ & 250 & 625 & 1250 & 1875 & 2500 & 7500 \\ \hline
  $L_1(\rho < 1)$ & 8.93$\times10^{-6}$ & 2.51$\times10^{-6}$ & 5.20$\times10^{-6}$ & 3.02$\times10^{-6}$ & 1.73$\times10^{-5}$ & 8.64$\times10^{-5}$ \\
  $L_2(\rho < 1)$ & 1.01$\times10^{-5}$ & 3.03$\times10^{-6}$ & 5.48$\times10^{-6}$ & 3.53$\times10^{-6}$ & 5.75$\times10^{-5}$ & 8.64$\times10^{-5}$ \\
  $L_{\infty}(\rho < 1)$ & 1.75$\times10^{-6}$ & 8.80$\times10^{-6}$ & 1.03$\times10^{-5}$ & 1.07$\times10^{-5}$ & 2.57$\times10^{-5}$ & 9.38$\times10^{-5}$ \\ \hline
  $L_1(1 < \rho < X)$ & 2.44$\times10^{-2}$ & 4.14$\times10^{-3}$ & 6.68$\times10^{-4}$ & 2.30$\times10^{-4}$ & 1.85$\times10^{-4}$ & 1.30$\times10^{-4}$ \\
  $L_2(1 < \rho < X)$ & 5.13$\times10^{-2}$ & 6.99$\times10^{-3}$ & 1.14$\times10^{-3}$ & 4.34$\times10^{-4}$ & 2.58$\times10^{-4}$ & 1.72$\times10^{-4}$ \\
  $L_{\infty}(1 < \rho < X)$ & 2.65$\times10^{-1}$ & 3.33$\times10^{-2}$ & 5.44$\times10^{-3}$ & 2.38$\times10^{-3}$ & 1.15$\times10^{-3}$ & 4.41$\times10^{-4}$ \\
  \hline
\end{tabular}    \caption{Mean error measures ($L_1$, $L_2$, and $L_{\infty}$) as a function of non-dimensional time $\tau$, inside and outside the cell. The calculations were based on the case $A=1/50$, $b=X=10$, and for FEM grid dimensions of $Q_1 \times Q_2 = 50 \times 20$ and a grid spacing of $h=0.2$.}
    \label{tab:table2}
\end{table}

Finally, to return to the first qualitative point raised at the beginning of this section, we investigate the effect of reducing the radius of our effective, virtual sphere boundary, $X$.

\begin{figure}[t!]
  \centering
  \includegraphics[width=0.3\columnwidth]{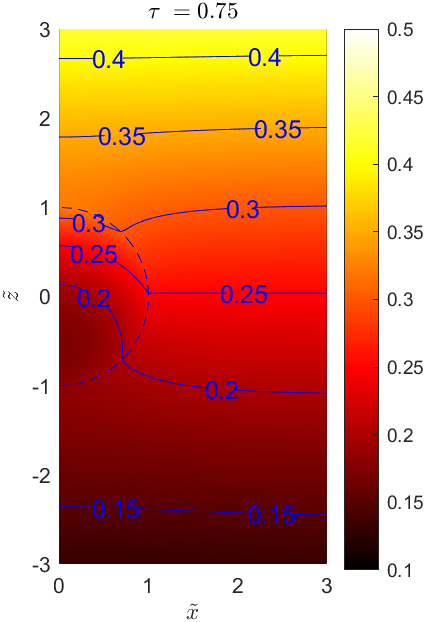}
  \hspace{0.15cm}
  \includegraphics[width=0.3\columnwidth]{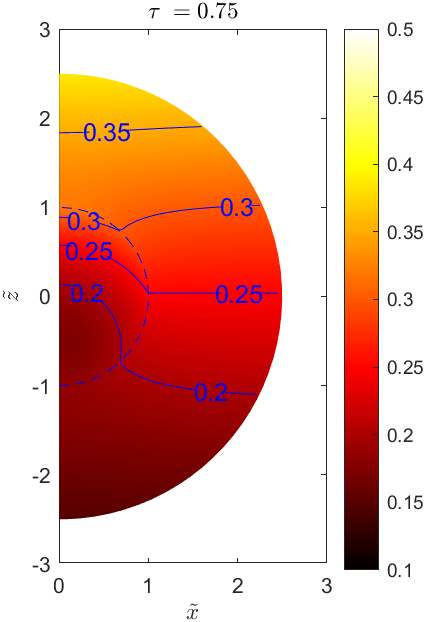}
  \hspace{0.15cm}
  \includegraphics[width=0.3\columnwidth]{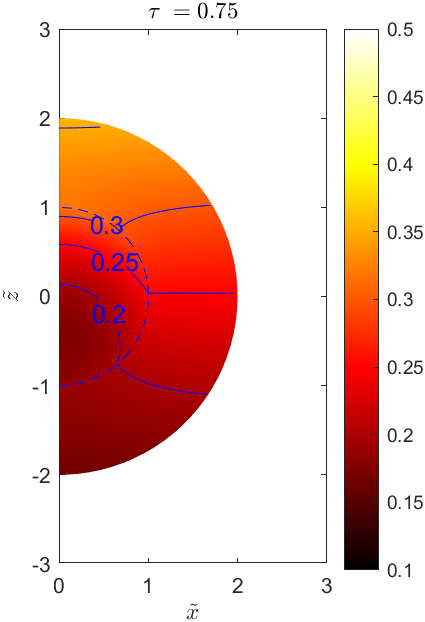}
  \hspace{0.15cm}
 \caption{The effect of reducing the radius of outer, virtual sphere boundary on the non-dimensional particle distributions inside and outside the biological cell predicted by the asymptotic approximation. We show the case for $A=50$, and $b=10$, at a fixed non-dimensional time of $\tau = 0.75$. The cases shown, from left to right, correspond to non-dimensional outer boundary radii of $X = 5$, 2.5, and 2. The most accurate asymptotic calculation, other things being equal, as well as the FEM benchmark result, is shown in Figure \ref{fig:figure4}.}
  \label{fig:figure6}
\end{figure}

Keeping fixed the distance to the physical source ($b=10$) the effect on the particle distribution at $\tau=0.75$ for $A=50$, of decreasing boundary radius, $X$ = 5, 2.5, 2 is demonstrated in Figure \ref{fig:figure6}. In theory, the most accurate example with this parameter set assumes $X=10$, this (along with the benchmark FEM results) was shown in Figure \ref{fig:figure4}. As we had chosen to maintain the figure scale for comparison purposes, the cases for which $X<3$ highlight the absence of perturbation solution data outside the virtual boundary. At the boundary, $\rho=X$, and beyond, the presupposition is for the particle distribution to be that of the unidirectional diffusion, Eq \eqref{OneDsoln}. A qualitative indicator of support for the validity of this assumption is a zero slope of the iso-contours there. From Figure \ref{fig:figure4} this qualitative support is present for $X=10$. Now in Figure \ref{fig:figure6} it is also apparently the case for $X=5$. However, for $X=2.5$ the iso-contours already show a nonzero slope at the virtual sphere boundary indicating that although the concentration values at $\rho=X$ are those of the unidirectional solution, there is a discontinuity in the slope in the $z$-direction. Consequently, for this radius at least there is a failure to merge smoothly with the unidirectional solution. Interestingly, we see from Table \ref{tab:table3}, that the accuracy of the asymptotic solution, relative to the FEM solution (which is independent of X), does not improve significantly for virtual sphere radii larger than $X=5$, indicating that it is not necessary to adopt the greatest possible sphere size to achieve reasonable accuracy.

\begin{table}
\centering
\begin{tabular}{c|c|c|c|c|c}
  \hline
  $X$ & 10 & 7.5 & 5 & 2.5 & 2 \\ \hline
  $m$ (exterior) & 4472 & 2495 & 1069 & 229 & 132 \\ \hline
  $L_1(\rho < 1)$ & 8.51$\times10^{-5}$ & 7.97$\times10^{-5}$ & 1.16$\times10^{-4}$ & 8.99$\times10^{-4}$ & 1.80$\times10^{-3}$ \\
  $L_2(\rho < 1)$ & 1.24$\times10^{-4}$ & 1.22$\times10^{-4}$ & 1.59$\times10^{-4}$ & 1.10$\times10^{-3}$ & 2.12$\times10^{-3}$ \\
  $L_{\infty}(\rho < 1)$ & 3.49$\times10^{-4}$ & 3.27$\times10^{-4}$ & 4.19$\times10^{-4}$ & 2.20$\times10^{-3}$ & 4.20$\times10^{-3}$ \\ \hline
  $L_1(1 < \rho < X)$ & 2.67$\times10^{-4}$ & 1.46$\times10^{-4}$ & 3.60$\times10^{-4}$ & 1.90$\times10^{-3}$ & 3.20$\times10^{-3}$ \\
  $L_2(1 < \rho < X)$ & 4.81$\times10^{-4}$ & 1.86$\times10^{-4}$ & 4.15$\times10^{-4}$ & 2.08$\times10^{-3}$ & 3.53$\times10^{-3}$ \\
  $L_{\infty}(1 < \rho < X)$ & 2.38$\times10^{-3}$ & 6.34$\times10^{-4}$ & 8.23$\times10^{-4}$ & 3.60$\times10^{-3}$ & 5.70$\times10^{-3}$ \\
  \hline
\end{tabular}    \caption{Mean error measures ($L_1$, $L_2$, and $L_{\infty}$) inside and outside the cell as a function of radius of outer, virtual spherical boundary $X$. The calculations were based on the case $A=50$, $b=10$, and at a non-dimensional time of $\tau=0.75$. The number of interior points sampled is fixed at $m=50$.}
    \label{tab:table3}
\end{table}

\section{Summary and Concluding Remarks} \label{sec:conc}

The asymptotic solution to the problem of particle diffusion in and around a spherical cell exposed to a plane source of particles is compared with a fully numerical solutions using the finite element method. We have indicated regimes in which the asymptotic solution is expected to perform poorly or fail altogether, as well as those regimes when it is expected to be accurate. The arguments have been confirmed by the numerical comparison with the FEM solutions. In fact, the asymptotic solution is shown to be surprisingly accurate over a considerable range of system parameters, even when the Legendre series is truncated at $n=5$ and the asymptotic series is truncated at $k=2$.

We found that good agreement, often to four decimal places, between the asymptotic approximation and the FEM solution occurred at considerably different non-dimensional times for different $A$ values. However, it was also found that these different non-dimensional times were in correspondence with the relative values of the diffusion constants, and agreement occurred at the same physical time.

The principal aim achieved in this work was the demonstration of good agreement between our asymptotic solution and a conventional, finite element solution of the diffusion equations. The agreement validates the use of the asymptotic solution expression (explicitly presented in \cite{Miklavcic2024}), instead of less convenient tables of numerical values, to determine derived quantities that are accessible experimentally. On this last note, an explicit example application is considered in \cite{Miklavcic2024b}.

An experimental detail, not yet explored theoretically but clearly warranted as indicated by the experimental studies, was the influence of gravity in the diffusion process. It was found by experiment~\cite{cui2016} that the orientation of the source-to-cell line relative to the direction of the gravitational force affected the rate at which nanoparticles accumulated in cells. Thus, it remains to include sedimentation in a future model. Such a theoretical analysis of the combined effect of sedimentation and diffusion would ascertain to what degree and in what form sedimentation affects the time dependent accumulation of nanoparticles in a spherical cell. We shall pursue this problem and will report on our findings, including a comparison with experimental data, in a separate publication.

\section*{Conflict of Interest Statement}
The authors declare that the research was conducted in the absence of any commercial or financial relationships that could be construed as a potential conflict of interest.

\section*{Author Contributions}
SJM was responsible for the model development, mathematical methodology, research direction, interpreting of results and drafting of the paper. SSK was responsible for independently verifying the analysis, undertaking the numerical study (FEM and asymptotic approximation), and preparing figures. Both authors reviewed and approved the final version of the manuscript.\par

\section*{Funding}
This project is supported by the Australian Research Council (Discovery project grant DP200103168).\par

\section*{Acknowledgments}

\bibliography{Kumar_Miklavcic_II_2024_arXiv}

\newpage

\appendix
\renewcommand{\theequation}{A.\arabic{equation}}.
\section{Separation of Variables Solution in the Laplace Domain}\label{sec: Separation of variables method}
The function $\overline{C}_2$,
\begin{equation}\label{eq:C_1 over bar i_n k_n P_n}
    \overline{C}_2=\sum_{n=0}^{\infty}(D_n i_n(\rho\sqrt{p})+E_n k_n(\rho\sqrt{p})) P_n(\cos(\phi)),
\end{equation}
describes the Laplace transformed concentration inside the cell.
Analogously, the function $\overline{C}_1$,
\begin{equation}\label{eq:C_2 over bar i_n k_n P_n}
    \overline{C}_1=\sum_{n=0}^{\infty}(A_n i_n(\rho\sqrt{p/A})+B_n k_n(\rho\sqrt{p/A})) P_n(\cos(\phi)),
\end{equation}
represents the Laplace transformed concentration outside the cell.
To fully determine these solutions we enforce the boundary conditions in the Laplace domain to determine the coefficients $A_n$, $B_n$, $D_n$ and $E_n$. For example, we observe that to eliminate the singularity at the origin $\rho = 0$ we require
\begin{equation}
    E_n=0, \quad n=0,1,2,\ldots \, ,
\end{equation}

Furthermore, to ensure continuity of concentration across the cell surface (at $\rho=1$) we require,
\begin{equation}\label{eq: LE concentration}
     A_n i_n(\sqrt{p/A})+B_n k_n(\sqrt{p/A})=D_n i_n(\sqrt{p}) \quad n=0,1,2,\ldots \, .
\end{equation}
Similarly, to ensure continuity of the flux density across the cell surface we require,
\begin{equation}
\sqrt{A}\Big(  A_n i_n'(\sqrt{p/A})+B_n k_n'(\sqrt{p/A})\Big) = D_n i_n'(\sqrt{p}) \label{eq: LE flux }
\end{equation}
where for brevity we define
\begin{equation}
i_n'(x)= \frac{\partial (i_n(x))}{\partial x}, \quad \quad k_n'(x)= \frac{\partial (k_n(x))}{\partial x}, \quad n=0,1,2, \ldots \, .
\end{equation}
Finally, we impose the far-field condition on the virtual spherical boundary, $X$. Utilising the expression known as the plane-wave expansion which can be found in~\cite{abram1965}(Eq. 10.1.47),  yields
\begin{equation}\label{eq: LE diffusive front }
    A_n i_n(X\sqrt{p/A})+B_n k_n(X \sqrt{p/A})=\frac{(2n+1)}{p}e^{-b\sqrt{p/A}}i_n(X \sqrt{p/A}), \quad n=0,1,\ldots \,.
\end{equation}

Equations~\ref{eq: LE concentration},~\ref{eq: LE flux } \&~\ref{eq: LE diffusive front } form a system of three linear equations in the unknown parameters $A_n$, $B_n$ and $D_n$.

To solve this system we first introduce the quantities,\footnote{Many of the following parameters and variables should additionally carry the index ``$n$''. However, in the interest of clarity we have suppressed the $n$. The reader should be mindful of the implicit dependence.}
 \begin{align}
    \sigma_l=i_n(x_l), \quad \omega_l&=k_n(x_l), \quad   \sigma_l'=\frac{\partial (i_n(x_l))}{\partial x_l}, \quad \omega_l'=\frac{\partial
    (k_n(x_l))}{\partial x_l}, \quad l=1,2,\ldots,5\label{eq:sigma_ omega_l}\\ \shortintertext{where}
    & \{x_1,x_2,\ldots, x_5\}=\{\sqrt{p},\, \rho \sqrt{p},\, \sqrt{\frac{p}{A}},\, \rho \sqrt{\frac{p}{A}},\, X\sqrt{\frac{p}{A}} \} \label{eq: x_l}.
\end{align}
 Solving the system of linear equations in terms of these new quantities, we obtain
\begin{align}
    A_n&=\Big(\frac{(2n+1)}{p}e^{-b\sqrt{\frac{p}{A}}}\Big) \frac{-(\sigma_1' \omega_3-\sqrt{A}\sigma_1 \omega_3')\sigma_5}{\omega_5(\sigma_1' \sigma_3-\sqrt{A}\sigma_1 \sigma_3')-\sigma_5(\sigma_1'\omega_3-\sqrt{A}\sigma_1\omega_3')},\\
    B_n&=\Big(\frac{(2n+1)}{p}e^{-b\sqrt{\frac{p}{A}}}\Big) \frac{(\sigma_1' \sigma_3-\sqrt{A}\sigma_1 \sigma_3')\sigma_5}{\omega_5(\sigma_1' \sigma_3-\sqrt{A}\sigma_1 \sigma_3')-\sigma_5(\sigma_1'\omega_3-\sqrt{A}\sigma_1\omega_3')},\\
    D_n&=\Big(\frac{(2n+1)}{p}e^{-b\sqrt{\frac{p}{A}}}\Big) \frac{\sqrt{A}(\sigma_3' \omega_3-\sigma_3 \omega_3')\sigma_5}{\omega_5(\sigma_1' \sigma_3-\sqrt{A}\sigma_1 \sigma_3')-\sigma_5(\sigma_1'\omega_3-\sqrt{A}\sigma_1\omega_3')}.
\end{align}
To summarize, the transformed solutions, $\overline{C}_1$ and $\overline{C}_2$, for the concentration outside and inside the cell, respectively, are
\begin{align}
   \overline{C}_2&=\sum_{n=0}^{\infty}\overline{V}_n P_n(\cos(\phi)),\label{eq: summary of sol 1} \\\shortintertext{where}
    \overline{V}_n&=\Big(\frac{(2n+1)}{p}e^{-b\sqrt{\frac{p}{A}}}\Big) \frac{\sqrt{A}(\sigma_3' \omega_3-\sigma_3 \omega_3')\sigma_5\sigma_2}{\omega_5(\sigma_1' \sigma_3-\sqrt{A}\sigma_1 \sigma_3')-\sigma_5(\sigma_1'\omega_3-\sqrt{A}\sigma_1\omega_3')}\label{eq: summary of sol 2}
\end{align}
and
\begin{align}
     \overline{C}_1&=\sum_{n=0}^{\infty}\overline{U}_n P_n(\cos(\phi)), \label{eq: summary of sol 3}\\\shortintertext{with}
     \overline{U}_n&=\Big(\frac{(2n+1)}{p}e^{-b\sqrt{\frac{p}{A}}}\Big) \frac{-(\sigma_1' \omega_3-\sqrt{A}\sigma_1 \omega_3')\sigma_5\sigma_4+(\sigma_1' \sigma_3-\sqrt{A}\sigma_1 \sigma_3')\sigma_5\omega_4}{\omega_5(\sigma_1' \sigma_3-\sqrt{A}\sigma_1 \sigma_3')-\sigma_5(\sigma_1'\omega_3-\sqrt{A}\sigma_1\omega_3')}.\label{eq: summary of sol 4}
\end{align}

\newpage

\renewcommand{\theequation}{B.\arabic{equation}}.
\section{Asymptotic Approximation}\label{sec: Asymptotic analysis}
The solutions in the Laplace domain given by Eqs.~\ref{eq: summary of sol 1}--\ref{eq: summary of sol 4} are complicated given the appearance of the Laplace parameter in the arguments of the modified spherical Bessel functions and in the majority of factors multiplying the Bessel functions. Consequently, achieving a closed form expression for the inverse Laplace transform is highly unlikely. Instead, as a compromise solution we proceed with an asymptotic analysis.

It is known that the Laplace transform $\overline{f}(p)$ of a function  as  $p \to 0$ corresponds to an approximation for $f(\tau)$ valid for $\tau \to \infty$.  This principle is discussed by Hahn and {\"O}zisik~\cite{HC:Book} and is used by Philip~\cite{philip1964} and Mild~\cite{mild1971}. A similar concept is formulated by Carslaw and Jaeger~\cite{OMIAM:Book}. We shall take advantage of this fact to obtain asymptotic approximations to the Laplace inverse of Eqs.~\ref{eq: summary of sol 1}--\ref{eq: summary of sol 4} valid for large times. 

\subsection{Large-time, asymptotic approximation for the concentration inside the cell}\label{sec: Large time asymptotic approximation inside the cell}
We begin by developing a series expansion of $\overline{C}_2$ as $p \to 0$. We rewrite Eqs.~\ref{eq: summary of sol 1}~\&~\ref{eq: summary of sol 2} in terms of our new quantities, Eqs.~\ref{eq:sigma_ omega_l} \& \ref{eq: x_l},
\begin{equation}\label{eq: overbar C_1 and overbar V_n}
    \overline{C}_2=\sum_{n=0}^{\infty}\overline{V}_n P_n(\cos(\phi)),\quad  \overline{V}_n=\Big(\frac{(2n+1)}{p}e^{-b\sqrt{\frac{p}{A}}}\Big)\frac{\Omega_1}{\Omega_2}
\end{equation}
where
\begin{equation}\label{eq: Omega_1}
    \Omega_1=\sqrt{A}(\sigma_3^{\prime }  \omega_3-\sigma_3 \omega_3')\sigma_5\sigma_2,
\end{equation}
\begin{equation}\label{eq: Omega_2}
    \Omega_2=\omega_5(\sigma_1' \sigma_3-\sqrt{A}\sigma_1 \sigma_3')-\sigma_5(\sigma_1'\omega_3-\sqrt{A}\sigma_1\omega_3').
\end{equation}
We first determine an expansion for $1/ \Omega_2$. We introduce the quantities
\begin{equation}
    \beta_l=i_{-n}(x_l),\quad  \beta_l'=\frac{\partial (i_{-n}(x_l))}{\partial x_l}, \quad l=1,2,\ldots,5
\end{equation}
where $i_{-n}(z)$ is the negative order modified spherical Bessel function of the first kind and with $x_l$ as defined in Eq.~\ref{eq: x_l}. Using Eq.~\ref{eq:i_n i_-n k_n} (in~\ref{sec: Modified spherical Bessel functions}), we have
\begin{equation}\label{eq: omega in terms of sigma and beta}
    \omega_l=\frac{(-1)^{n+1}\pi}{2}(\sigma_l-\beta_l)
\end{equation}
and
\begin{equation}\label{eq: omega' in terms of sigma' and beta'}
    \omega_l'=\frac{(-1)^{n+1}\pi}{2}(\sigma_l'-\beta_l').
\end{equation}
We substitute Eqs.~\ref{eq: omega in terms of sigma and beta} \& \ref{eq: omega' in terms of sigma' and beta'} into Eq.~\ref{eq: Omega_2}, to obtain
\begin{equation}\label{eq: Omega2 in terms of sigmas and betas}
    \Omega_2=\frac{(-1)^{n+1}\pi}{2 }(\beta_5(\sqrt{A}\sigma_1\sigma_3'-\sigma_1'\sigma_3)-\sigma_5(\sqrt{A}\sigma_1\beta_3'-\sigma_1'\beta_3)).
\end{equation}
We now introduce an expansion for each $\sigma_l$, which are the modified spherical Bessel functions of the first kind, using Eq.~\ref{eq: expansion of i_n i_-n}:
\begin{equation}\label{eq: expansion in terms of sigma}
\sigma_l=(\sqrt{p})^{n}\sum_{k=0}^{\infty}\sigma_{l,k}\, (\sqrt{p})^{2k}
\end{equation}
where
\begin{equation}\label{eq: definition of sigma_l,k}
 \sigma_{l,k}=\frac{a(n,k)(x_l)^{2k+n}}{(\sqrt{p})^{2k+n}},\qquad k=0,1,\ldots\,,
\end{equation}
$a(n,k)$,  defined  in Eq.~\ref{eq: a(n,k) b(n,k)} are coefficients that are independent of $p$.
We note that $\sigma_{l,k}$ are the coefficients of $(\sqrt{p})^{2k+n}$ in the expansion for $i_n(x_l)$ using Equation~\ref{eq: expansion of i_n i_-n}. We observe that $x_l$ (Equation~\ref{eq: x_l}) is comprised of $\sqrt{p}$  in the numerator, for each $l$. The factor of $\sqrt{p}$ is eliminated from  $\sigma_{l,k}$ by the  $(\sqrt{p})^{2k+n}$ in the denominator of Equation~\ref{eq: definition of sigma_l,k}.    Hence,  $\sigma_{l,k}$ for all values of $k$ and $l$ are independent of $p$. Similarly, we introduce expansions for $\sigma_l'$, $\beta_l$ and $\beta_l'$ using Equations~\ref{eq: expansion of i_n i_-n},~\ref{eq: expansion for i_n'} \&~\ref{eq: expansion for i_-n'}
\begin{equation}\label{eq: expansions in terms of sigma' beta}
\sigma'_l=(\sqrt{p})^{(n-1)}\sum_{k=0}^{\infty}\sigma'_{l,k}\, (\sqrt{p})^{2k}, \quad \beta_l=(\sqrt{p})^{-(n+1)} \sum_{k=0}^{\infty}\beta_{l,k}\, (\sqrt{p})^{2k},
\end{equation}
\begin{equation}\label{eq: expansions in terms of beta'}
 \beta_l'=(\sqrt{p})^{-(n+2)} \sum_{k=0}^{\infty}\beta_{l,k}\, (\sqrt{p})^{2k},  \quad l=1,2,\ldots,5
\end{equation}
where
\begin{equation}\label{eq: definition of sigma_l,k' beta_l,k}
\sigma_{l,k}'=\frac{a(n,k)(2k+n)(x_l)^{2k+(n-1)}}{(\sqrt{p})^{2k+(n-1)}},\quad \beta_{l,k}=\frac{b(n,k)(x_l)^{2k-(n+1)}}{(\sqrt{p})^{2k-(n+1)}}, \quad \end{equation}
\begin{equation}\label{eq: definition of beta_l,k'}
\beta_{l,k}'=\frac{b(n,k)(2k-n-1)(x_l)^{2k-(n+2)}}{(\sqrt{p})^{2k-(n+2)}}, \quad k=0,1,...\, .
\end{equation}
Analogous to $\sigma_{l,k}$, we observe that $\sigma_{l,k}'$, $\beta_{l,k}$ and $\beta_{l,k}'$ are not dependent on $p$ for all values of $l$ and $k$. Substituting Eqs.~\ref{eq: expansion in terms of sigma},~\ref{eq: expansions in terms of sigma' beta} \& \ref{eq: expansions in terms of beta'} into Eq.~\ref{eq: Omega2 in terms of sigmas and betas}, we obtain
\begin{equation}
    \Omega_2= \frac{(-1)^{n+1}\pi \, p^{n/2-1}}{2 }(\Phi_1-\Phi_2)
\end{equation}
where
\begin{align}
\Phi_1&=    \Big(\sum_{k=0}^{\infty}\beta_{5,k}\, p^k \Big) \Bigg(\sqrt{A}\Big(\sum_{k=0}^{\infty}\sigma_{1,k}\, p^k \Big)\Big(\sum_{k=0}^{\infty}\sigma_{3,k}'\, p^k \Big)-\Big(\sum_{k=0}^{\infty}\sigma_{1,k}'\, p^k \Big)\Big(\sum_{k=0}^{\infty}\sigma_{3,k}\, p^k \Big)\Bigg),\\
\Phi_2&=    \Big(\sum_{k=0}^{\infty}\sigma_{5,k}\, p^k \Big) \Bigg(\sqrt{A}\Big(\sum_{k=0}^{\infty}\sigma_{1,k}\, p^k \Big)\Big(\sum_{k=0}^{\infty}\beta_{3,k}'\, p^k \Big)-\Big(\sum_{k=0}^{\infty}\sigma_{1,k}'\, p^k \Big)\Big(\sum_{k=0}^{\infty}\beta_{3,k}\, p^k \Big)\Bigg).
\end{align}
The expressions for $\Phi_1$ and $\Phi_2$ are composed of terms which are products of three series in $p$.  Using the triple Cauchy product formula from Eq.~\ref{eq: triple cauchy product}, we get the single series for the denominator in Eq.~\ref{eq: overbar C_1 and overbar V_n}
\begin{equation}\label{eq: Omega2 exapsnion in terms of gamma}
    \Omega_2=p^{n/2-1} \sum_{k=0}^{\infty} \gamma_k \, p^k
\end{equation}
where
\begin{equation}\label{eq: definition of gamma_k}
    \gamma_k=\frac{(-1)^{n+1}\pi}{2 }
    \sum_{i=0}^{k} \sum_{j=0}^{i} \beta_{5,k-i} (\sqrt{A} \sigma_{1,i-j}\, \sigma_{3,j}' -  \sigma_{1,i-j}'\, \sigma_{3,j})- \sigma_{5,k-i}(\sqrt{A}\sigma_{1,i-j}\, \beta_{3,j}'-\sigma_{1,i-j}' \,\beta_{3,j}).
\end{equation}
Manipulating Eq.~\ref{eq: Omega2 exapsnion in terms of gamma}, we obtain the expression
\begin{equation}\label{eq: 1/Omega_2}
    \frac{1}{\Omega_2}= \frac{ p^{(1-n/2)}}{\gamma_0}\Big(1+ \sum_{k=1}^{\infty}\frac{ \gamma_k}{\gamma_0} \, p^k\Big)^{-1}
\end{equation}
Using the geometric series expansion~\cite{HC:Book}, the last equation may be re-written as 
\begin{equation}\label{eq: Omega2 final expansion}
    \frac{1}{\Omega_2}= p^{(1-n/2)}\sum_{r=0}^{\infty} \frac{(-1)^r}{\gamma_0}\Big(\sum_{k=1}^{\infty}\frac{ \gamma_k}{\gamma_0} \, p^k\Big)^r=p^{(1-n/2)}\sum_{k=0}^{\infty} \mu_k\, p^k,
\end{equation}
which is valid for small values of $p$ since
\begin{equation}
    \Big|\sum_{k=1}^{\infty}\frac{ \gamma_k}{\gamma_0} p^k \Big|<1, \qquad\textrm{as} \qquad p \to 0.
\end{equation}
We determine the first four terms of $\mu_k$,
\begin{equation}\label{eq: mu gamma relation}
 \mu_0=\frac{1}{\gamma_0},\quad
 \mu_1=-\frac{\gamma_1}{\gamma_0^2},\quad
 \mu_2=\frac{\gamma_1^2}{\gamma_0^3}-\frac{\gamma_2}{\gamma_0^2}, \quad \mu_3=-\frac{\gamma_1^3}{\gamma_0^4}+2\frac{\gamma_1 \gamma_2}{\gamma_0^3}-\frac{\gamma_3}{\gamma_0^2}\, .
\end{equation}
We now develop a series expansion for $\Omega_1$ in the numerator of Eq.~\ref{eq: overbar C_1 and overbar V_n}. We observe that the term
$\sigma_3' \omega_3-\sigma_3 \omega_3'$ in Eq.~\ref{eq: Omega_1}
is the Wronskian of $i_n(x_3)$ and $k_n(x_3)$ with $x_3=\sqrt{p/A}$. Hence, using Eq.~\ref{eq: Wronskian}, we find
\begin{equation}
    \sigma_3' \omega_3-\sigma_3 \omega_3'=-\frac{\pi A}{2p}.
\end{equation}
Substitution of this into Eq.~\ref{eq: Omega_1}, we get
\begin{equation}
    \Omega_1=-\frac{A^{3/2} \pi }{2p} \sigma_5 \sigma_2.
\end{equation}
Using Eq.~\ref{eq: expansion in terms of sigma}, we obtain
\begin{equation}\label{eq: Omega1 in terms of sigma}
    \Omega_1=-\frac{A^{3/2} \pi }{2}p^{n-1}\Big(\sum_{k=0}^{\infty}\sigma_{5,k}\,p^k  \Big) \Big( \sum_{k=0}^{\infty}\sigma_{2,k} \,p^k\Big).
\end{equation}
We substitute Eqs.~\ref{eq: Omega2 final expansion} \&~\ref{eq: Omega1 in terms of sigma} into~\ref{eq: overbar C_1 and overbar V_n}, to get our Legendre coefficient functions,
\begin{equation}
 \overline{V}_n=-\frac{(2n+1)A^{3/2}\pi}{2}e^{-b\sqrt{\frac{p}{A}}}p^{n/2-1}\Big( \sum_{k=0}^{\infty} \mu_k\, p^k\Big) \Big(\sum_{k=0}^{\infty}\sigma_{5,k}\,p^k  \Big) \Big( \sum_{k=0}^{\infty}\sigma_{2,k}\,p^k \Big).
\end{equation}
Using the  triple Cauchy product formula Eq.~\ref{eq: triple cauchy product}, we arrive at the final series expansion
\begin{equation}\label{eq: overline V_n expansion}
    \overline{V}_n=e^{-b\sqrt{\frac{p}{A}}}p^{n/2-1}\sum_{k=0}^{\infty} \psi_k^{(n)}\,p^k,
\end{equation}
where
\begin{equation}\label{eq: definition od psi_k}
    \psi_k^{(n)}=-\frac{(2n+1)A^{3/2}\pi}{2} \sum_{i=0}^{k} \sum_{j=0}^{i}  \mu_{k-i}\, \sigma_{5,i-j}\, \sigma_{2,j}, \quad n=0,1,2\ldots \, .
\end{equation}
We note that the radial dependence is contained in the factor $\sigma_{2,j}$, but more importantly, none of these factors contain $p$.
We now proceed to determine the inverse Laplace transform of $\overline{C}_2$, using Eqs.~\ref{eq: overbar C_1 and overbar V_n}~\&~\ref{eq: overline V_n expansion},
\begin{equation}
    \mathcal{L}^{-1}\{\overline{C}_2\}=\sum_{n=0}^{\infty}\mathcal{L}^{-1}\{\overline{V}_n\} P_n(\cos(\phi)).
\end{equation}
where
\begin{equation}
    \mathcal{L}^{-1}\{\overline{V}_n\}=\sum_{k=0}^{\infty} \psi_k^{(n)}\,\mathcal{L}^{-1}\{e^{-b\sqrt{\frac{p}{A}}}p^{k+n/2-1}\}\, .
\end{equation}
The inverse Laplace transform of an arbitrary term in the series, $e^{-b\sqrt{\frac{p}{A}}}p^{k+n/2-1}$, can now  be determined for all values of $n$ and $k$ using Eq.~\ref{eq: inverse laplace of e^(-1 sqrt p) p^n}. Performing the said inverse we arrive at our sought after asymptotic formula,
\begin{equation}\label{eq: C_1 with g(x,m,tau)}
    C_2=\sum_{n=0}^{\infty}\sum_{k=0}^{\infty} \psi_k^{(n)}\,g(-b/\sqrt{A},k+n/2-1,\tau) P_n(\cos(\phi))
\end{equation}
where
\begin{align}
&  g(x,m,\tau)= \nonumber \\
&\; \left\lbrace
\begin{array}{ll}
\tau^{-m} \left(\dfrac{\, _1F_1\left(m+1;\frac{1}{2};-\frac{x^2}{4 \tau}\right)}{\tau \Gamma (-m)}-\dfrac{x \,\,  _1F_1\left(m+\frac{3}{2};\frac{3}{2};-\frac{x^2}{4 \tau}\right)}{\tau^{3/2} \Gamma \left(-m-\frac{1}{2}\right)}\right) \quad & \;\textrm{if}\;\;m\le -1,\\
-\dfrac{x\, \tau^{-m-3/2}  \, _1F_1\left(m+\frac{3}{2};\frac{3}{2};-\frac{x^2}{4 \tau}\right)}{ \Gamma \left(-m-\frac{1}{2}\right)}& \;\textrm{if}\;\;m=0,1,2,\ldots \, ,\\
 \dfrac{\,\tau^{-m-1} \, _1F_1\left(m+1;\frac{1}{2};-\frac{x^2}{4 \tau}\right)}{ \Gamma (-m)}& \;\textrm{if}\;\; m=i+1/2,\; i=-1,0,1,2,\ldots\, ,\label{eq: definition of G(x,m,tau)}
\end{array}\right.
\end{align}
and $_1F_1$ is the confluent hypergeometric function of the first kind~\cite{mf1953}.

The explicit forms of the $\psi_k^{(n)}$ are very large due to the consecutive Cauchy products required for its computation. In the remainder of this section we consider an approximation to Equation~\ref{eq: C_1 with g(x,m,tau)}, by considering just the $k=0$ term of the inner summation. Hence, we get
\begin{equation}
  C_2\approx\sum_{n=0}^{\infty} \psi_0^{(n)}\,g(-b/\sqrt{A},n/2-1,\tau) P_n(\cos(\phi))  \, .
\end{equation}
Using Eqs. \ref{eq: definition of sigma_l,k}, \ref{eq: definition of sigma_l,k' beta_l,k}, \ref{eq: definition of beta_l,k'}, \ref{eq: definition of gamma_k}, \ref{eq: mu gamma relation}, \ref{eq: definition od psi_k} \& \ref{eq: a(n,k) b(n,k)}, we obtain
\begin{equation}
  \psi_0^{(n)}=  \frac{\sqrt{\pi}\,2^{-n} \,A^{1-\frac{n}{2}}  \,X^{1+2\,n} \,\rho^n \,{\left(1+2\,n\right)}}{{\Gamma }\left(\frac{1}{2}+n\right)\,{\left(A\,X^{1+2\,n} +n\,{\left(A+A\,X^{1+2\,n} +X^{1+2\,n} -1\right)}\right)}}.
\end{equation}

In Section~\ref{sec:numasymp} we examine a better approximation of Eq.~\ref{eq: C_1 with g(x,m,tau)} by considering $\psi_k^{(n)}$ for $k=0,1,2$.

\subsection{Large-time, asymptotic approximation for the concentration outside the cell}\label{sec: Large time asymptotic approximation  outside the cell}
We find the expansion for $\overline{C}_1$ as $p \to 0$ in a similar way to the procedure followed in~\ref{sec: Large time asymptotic approximation inside the cell}. We write Eqs.~\ref{eq: summary of sol 3} \& \ref{eq: summary of sol 4} using new quantities,
\begin{equation}\label{eq: over bar C2 interms of Omega2 and Omega3}
    \overline{C}_1=\sum_{n=0}^{\infty}\overline{U}_n P_n(\cos(\phi)), \quad
     \overline{U}_n=\Big(\frac{(2n+1)}{p}e^{-b\sqrt{\frac{p}{A}}}\Big) \frac{\Omega_3}{\Omega_2}
\end{equation}
where
\begin{equation}\label{eq: Omega_3}
    \Omega_3=-(\sigma_1' \omega_3-\sqrt{A}\sigma_1 \omega_3')\sigma_5\sigma_4+(\sigma_1' \sigma_3-\sqrt{A}\sigma_1 \sigma_3')\sigma_5\omega_4, \quad
\end{equation}
and where $\Omega_2$ is defined by Eq.~\ref{eq: Omega_2}.
To determine an expansion for $\Omega_3$, we first substitute Eqs.~\ref{eq: omega in terms of sigma and beta} \&~\ref{eq: omega' in terms of sigma' and beta'} into the above expression of $\Omega_3$ to get,
\begin{equation}
    \Omega_3=\frac{(-1)^{n+1}\pi}{2 }((\sigma_1'  \beta_3 -\sqrt{A} \sigma_1 \beta_3')\sigma_4  -(\sigma_1'\sigma_3-\sqrt{A}\sigma_1\, \sigma_3 )\beta_4) \sigma_5 \, .
\end{equation}
Then using the expansions defined by Eqs.~\ref{eq: expansion in terms of sigma},~\ref{eq: expansions in terms of sigma' beta} \& \ref{eq: expansions in terms of beta'}, and evaluating the product of these expansions  using the quadruple Cauchy product (Eq.~\ref{eq: quadruple cauchy product}), we get
\begin{equation}\label{eq: Omega_3 expansion}
    \Omega_3= p^{n-1}\sum_{k=0}^\infty \alpha_k \, p^k
\end{equation}
where
\begin{align}
    &\alpha_k=\frac{(-1)^{n+1}\pi}{2} \times \nonumber\\
    & \;  \sum_{i=0}^{k}\sum_{j=0}^i \sum_{l=0}^j \sigma_{5,k-i} ((\sigma_{1,i-j}'  \beta_{3,j-l} - \sqrt{A} \sigma_{1,i- j} \beta_{3,j-l}')\sigma_{4,l}  - (\sigma_{1,i-j}'\sigma_{3, j -l} - \sqrt{A}\sigma_{1, i - j}\, \sigma_{3, j - l} )\beta_{4, l}), \label{eq: definition of alpha_k}
\end{align}
where $\sigma_{l,k}$, $\sigma_{l,k}'$, $\beta_{l,k}$ and $\beta_{l,k}'$ are defined by Eqs.~\ref{eq: definition of sigma_l,k}, \ref{eq: definition of sigma_l,k' beta_l,k} \& \ref{eq: definition of beta_l,k'}.
Substituting Eqs.~\ref{eq: Omega2 final expansion} \& \ref{eq: Omega_3 expansion}  into Eq.~\ref{eq: over bar C2 interms of Omega2 and Omega3}, we have
\begin{equation}\label{eq: overbar U in terms of alpha and mu}
    \overline{U}_n=(2n+1) p^{n/2-1}e^{-b\sqrt{\frac{p}{A}}}\Big(\sum_{k=0}^\infty \alpha_k \, p^k \Big)\Big( \sum_{k=0}^{\infty} \mu_k\, p^k\Big)
\end{equation}
where $\mu_k$, again appears, defined by Eq.~\ref{eq: Omega2 final expansion}, the first four terms of which are explicitly stated in Eq.~\ref{eq: mu gamma relation}. Evaluating  Eq.~\ref{eq: overbar U in terms of alpha and mu} using the Cauchy product,
\begin{equation}\label{eq: overbar U in terms of nu}
    \overline{U}_n= p^{n/2-1}e^{-b\sqrt{\frac{p}{A}}} \sum_{k=0}^{\infty} \nu_k^{(n)} \, p_k
\end{equation}
where the coefficients
\begin{equation}\label{eq: definition of nu_k}
    \nu_k^{(n)}= (2n+1)  \sum_{i=0}^{k} \alpha_{k-i}\, \mu_i \, ,
\end{equation}
are independent of $p$. We are now in a position to determine the inverse Laplace transform of $\overline{C}_1$ using Eq.~\ref{eq: overbar U in terms of nu}. From Eq.~\ref{eq: over bar C2 interms of Omega2 and Omega3}, we have
\begin{equation}
    \mathcal{L}^{-1}\{\overline{C}_1\}=\sum_{n=0}^{\infty}\mathcal{L}^{-1}\{\overline{U}_n\} P_n(\cos(\phi)).
\end{equation}
where
\begin{equation}
    \mathcal{L}^{-1}\{\overline{U}_n\}=\sum_{k=0}^{\infty} \nu_k^{(n)}\,\mathcal{L}^{-1}\{e^{-b\sqrt{\frac{p}{A}}}p^{k+n/2-1}\}\, .
\end{equation}
The inverse Laplace transform of $e^{-b\sqrt{\frac{p}{A}}}p^{k+n/2-1}$ can be determined for all values of $n$ and $k$ using Eq.~\ref{eq: inverse laplace of e^(-1 sqrt p) p^n}. Hence we have

\begin{equation}\label{eq: C_2 with g(x,m,tau)}
    C_1=\sum_{n=0}^{\infty}\sum_{k=0}^{\infty} \nu_k^{(n)}\,g(-b/\sqrt{A},k+n/2-1,\tau) P_n(\cos(\phi))
\end{equation}
where $g(x,m,\tau)$ is again defined by Eq.~\ref{eq: definition of G(x,m,tau)}. The explicit forms of the series coefficients, $\nu_k^{(n)}$, for arbitrary $k$ and $n$ are very large due the consecutive Cauchy products required  in the computation. To conclude this section, however, we may consider a simple approximation to Equation~\ref{eq: C_2 with g(x,m,tau)} by considering just the $k=0$ term for the inner summation. Hence, we get
\begin{equation}
  C_1\approx\sum_{n=0}^{\infty} \nu_0^{(n)}\,g(-b/\sqrt{A},n/2-1,\tau) P_n(\cos(\phi))  \,.
\end{equation}
Using Eqs.~\ref{eq: definition of sigma_l,k}, \ref{eq: definition of sigma_l,k' beta_l,k}, \ref{eq: definition of beta_l,k'}, \ref{eq: definition of gamma_k}, \ref{eq: mu gamma relation}, \ref{eq: definition of alpha_k}, \ref{eq: definition of nu_k} \& \ref{eq: a(n,k) b(n,k)}, we find that
\begin{equation}
    \nu_0^{(n)}=\frac{\sqrt{\pi}\,2^{-1-n}  \,X^{1+2\,n} \,{\left(n\,{\left(A+A\,\rho^{1+2\,n} +\rho^{1+2\,n} -1\right)}+A\,\rho^{1+2\,n} \right)}\,{\left(1+2\,n\right)}}{A^{n/2} \,\rho^{1+n} \,{\Gamma }\left(\frac{3}{2}+n\right)\,{\left(A\,X^{1+2\,n} +n\,{\left(A+A\,X^{1+2\,n} +X^{1+2\,n} -1\right)}\right)}}\, .
\end{equation}

In Section~\ref{sec:numasymp} we examine a better approximation of Eq.~\ref{eq: C_2 with g(x,m,tau)} by considering $\nu_k^{(n)}$ for $k=0,1,2$.

\newpage

\renewcommand{\theequation}{C.\arabic{equation}}.
\section{Further Required Results}
\subsection{Modified spherical Bessel functions}\label{sec: Modified spherical Bessel functions}
 The modified spherical Bessel functions of the first and second kind are   $i_n(z)$ ( $i_{-n}(z)$ is of negative order) and $k_n(z)$. We use the definitions used by Olver et al. \cite{NIST:DLMF}, 
\begin{equation}\label{eq: definition od in i-n kn}
   i_n(z)=\sqrt{\frac{\pi}{2z}} I_{n+1/2}(z), \qquad i_{-n}(z)=\sqrt{\frac{\pi}{2z}} I_{-n-1/2}(z), \qquad  k_n(z)=\sqrt{\frac{\pi}{2z}} K_{n+1/2}(z),
\end{equation}
where $I_m(z)$ and $K_m(z)$ are the modified Bessel functions of the first and second kind of order $m$. We can relate  $k_n(z)$ to $i_n(z)$ and  $i_{-n}(z)$ using the  formula~\cite{NIST:DLMF} (Eq.~10.47.11), 
\begin{equation}\label{eq:i_n i_-n k_n}
    k_n(z)=\frac{(-1)^{n+1}\pi}{2}(i_n(z)-i_{-n}(z)).
\end{equation}

The Wronskian of $i_n(z)$ and $k_n(z)$ is known to be~\cite{NIST:DLMF} (Eq.~10.50.2), 
\begin{equation}\label{eq: Wronskian}
    i_n'(z)k_n(z)-i_n(z)k_n'(z)=-\frac{\pi}{2z^2}  \, .
\end{equation}

\subsection{Series expansions}
The series expansions of $i_n(z)$ and  $i_{-n}(z)$  in~\cite{NIST:DLMF} (Eq.~10.53.3 \& 10.53.4) are here denoted 
\begin{equation}\label{eq: expansion of i_n i_-n}
    i_n(z)=z^n\sum^{\infty}_{k=0} a(n,k) z^{2k}, \qquad   i_{-n}(z)=z^{-n-1}\sum^{\infty}_{k=0} b(n,k) z^{2k}
\end{equation}
where
\begin{equation}\label{eq: a(n,k) b(n,k)}
    a(n,k)=\frac{1}{2^k k!(2n+2k+1)!!} , \quad
b(n,k) = \left\{
        \begin{array}{ll}
            \dfrac{(-1)^{n+k}(2n-2k-1)!!}{2^k k!},
            & \quad \text{for} \quad k \leq n, \\
            \dfrac{1}{2^k k!(2k-2n-1)!!} ,
            & \quad \text{for} \quad k > n.
        \end{array}
    \right.
\end{equation}

Similarly, we have expressions of $i_n'(z)$ and $i_{-n}'(z)$,
\begin{align}
  & i_n'(z)=\frac{\partial(i_n(z))}{\partial z}& &i_n'(z)=z^{n-1}\sum^{\infty}_{k=0}(2k+n) a(n,k) z^{2k}, \label{eq: expansion for i_n'} \\
   & i_{-n}'(z)=\frac{\partial(i_{-n}(z))}{\partial z} & &i_{-n}'(z)=z^{-n-2}\sum^{\infty}_{k=0} (2k-n-1)b(n,k) z^{2k},\label{eq: expansion for i_-n'}
\end{align}
where the coefficients $a(n,k)$ and $b(n,k)$ are defined by Eq.~\ref{eq: a(n,k) b(n,k)}.

\subsection{Multiple Cauchy products}
We determine the product of three series.  Let,
\begin{equation}
 P_3= \Big( \sum^{\infty}_{k=0} a_k\, z^k\Big) \Big( \sum^{\infty}_{k=0} b_k\,z^k\Big) \Big( \sum^{\infty}_{k=0} c_k\, z^k\Big).
\end{equation}
Considering just the last two series,
\begin{equation}\label{eq: e_i}
       P_3= \Big( \sum^{\infty}_{k=0}a_k\, z^k\Big)\Big( \sum^{\infty}_{k=0} e_k\, z^k\Big), \quad  e_k=\sum^{k}_{j=0}b_{k-j}\, c_j \,,
\end{equation}
where  $e_k$ was determined by the usual  Cauchy product.
Using this operation again, we get
\begin{equation}
    P_3=\sum^{\infty}_{k=0}\Big(\sum^{k}_{i=0}a_{k-i}\, e_i \Big)\,z^k.
\end{equation}
Substituting Eq.~\ref{eq: e_i},  we obtain the triple Cauchy product formula
\begin{equation}\label{eq: triple cauchy product}
    P_3=\sum^{\infty}_{k=0} \Big( \sum^{k}_{i=0}\sum^{i}_{j=0}a_{k-i}\,  b_{i-j}\, c_j\Big) \,z^k \, .
\end{equation}
 Similarly, for the case of a product of four series, we  have the quadruple Cauchy product formula,
\begin{equation}\label{eq: quadruple cauchy product}
    \Big( \sum^{\infty}_{k=0} a_k z^k \Big) \Big( \sum^{\infty}_{k=0} b_k z^k \Big) \Big( \sum^{\infty}_{k=0}  c_k z^k \Big)\Big(\sum^{\infty}_{k=0}  d_k z^k \Big)=\sum_{k=0}^{\infty} \Big(
    \sum_{i=0}^{k}\sum_{j=0}^i \sum_{l=0}^j a_{k-i}\, b_{i-j}\, c_{j-l}\, d_l\Big)\,z^k \,.
\end{equation}

\subsection{Laplace inverse of expansions}
 Equation~\ref{eq: inverse laplace of e^(-1 sqrt p) p^n} below is used to document the inverse Laplace transform of the expansions developed in \ref{sec: Asymptotic analysis}.~We~have,
\begin{align}
&  \mathcal{L}^{-1}\{e^{-x\sqrt{p}}p^m\}= \nonumber \\
&\; \left\lbrace
\begin{array}{ll}
\tau^{-m} \left(\dfrac{\, _1F_1\left(m+1;\frac{1}{2};-\frac{x^2}{4 \tau}\right)}{\tau \Gamma (-m)}-\dfrac{x \, _1F_1\left(m+\frac{3}{2};\frac{3}{2};-\frac{x^2}{4 \tau}\right)}{\tau^{3/2} \Gamma \left(-m-\frac{1}{2}\right)}\right) \quad & \;\textrm{if}\;\;m\le -1,\\
-\dfrac{x\, \tau^{-m-3/2}  \, _1F_1\left(m+\frac{3}{2};\frac{3}{2};-\frac{x^2}{4 \tau}\right)}{ \Gamma \left(-m-\frac{1}{2}\right)}& \;\textrm{if}\;\;m=0,1,2,\ldots \, ,\\
 \dfrac{\,\tau^{-m-1}\, _1F_1\left(m+1;\frac{1}{2};-\frac{x^2}{4 \tau}\right)}{ \Gamma (-m)}& \;\textrm{if}\;\; m=i+1/2,\; i=-1,0,1,2,\ldots\, ,
\end{array}\right.\label{eq: inverse laplace of e^(-1 sqrt p) p^n}
\end{align}
where
\begin{equation}\label{eq:  confluent hypergeometric function of the first kind}
    _1F_1(a;b;z)=\sum_{k=0}^\infty \frac{\Gamma(a+k)\Gamma(b)}{\Gamma(b+k)\Gamma(a)\Gamma(k+1)}z^k,
\end{equation}
$x>0$, $p$ is the Laplace parameter and $_1F_1$ is the confluent hypergeometric function of the first kind~\cite{CHFOFKMathWorld}.  Equation~\ref{eq: inverse laplace of e^(-1 sqrt p) p^n} was determined using the function \verb|InverseLaplaceTransform| by Wolfram Mathematica~\cite{ILTFunctionWM}.  The expression condenses to a more comprehensible expressions for specific values of $m$. For instance,  $m=-1$
\begin{equation}
     \mathcal{L}^{-1}\{e^{-x\sqrt{p}}p^{-1}\}=\, _1F_1\left(0;\frac{1}{2};-\frac{x^2}{4 \tau}\right)-\frac{x \,\, _1F_1\left(\frac{1}{2};\frac{3}{2};-\frac{x^2}{4 \tau}\right)}{\sqrt{\tau\pi}}.
\end{equation}
Using formulas in~\cite{NIST:DLMF} (Eq.~13.6.3 \&~13.6.7) we observe 
\begin{equation}
 \mathcal{L}^{-1}\{e^{-x\sqrt{p}}p^{-1}\}=\text{erfc}\left(\frac{x}{2 \sqrt{\tau}}\right) ,
\end{equation}
which agrees with the expressions in~\cite{crank1956,HC:Book}.

\end{document}